\documentclass[superscriptaddress,twocolumn,10pt,prx,aps,showpacs,longbibliography,floatfix]{revtex4-2}
\usepackage{graphicx}
\usepackage[dvipsnames]{xcolor}
\usepackage{physics}
\usepackage{dcolumn}
\usepackage{bm}
\usepackage{comment}
\usepackage{siunitx}
\usepackage{mathtools}

\usepackage{float}
\usepackage{array}
\usepackage{amsbsy}
\usepackage[utf8x]{inputenc}
\usepackage{epstopdf}
\usepackage{amsmath,amssymb,amsfonts}
\usepackage{booktabs}
\usepackage{dsfont}
\usepackage{cprotect}
\usepackage{mathtools}
\usepackage[normalem]{ulem}
\usepackage{multirow}
\usepackage{verbatim}
\usepackage{tabularx}
\usepackage{enumitem}
\usepackage{url}
\usepackage[colorlinks]{hyperref}
\hypersetup{
    plainpages=true,
    breaklinks=true,
    hypertexnames=false,
    pageanchor=true,
    colorlinks=true,
    linkcolor={blue},
    citecolor={red},
    urlcolor={blue},
    anchorcolor={black}
}

\usepackage[colorlinks]{hyperref}
\hypersetup{
	plainpages=true,
	breaklinks=true,
	hypertexnames=false,
	pageanchor=true,
	colorlinks=true,
	linkcolor={blue},
	citecolor={red},
	urlcolor={blue},
	anchorcolor={black}
}

\newcommand{\LL}[0]{\mathcal{L}}

\newcommand{\D}[0]{\mathcal{D}}
\newcommand{\ad}[0]{{\hat{a}^\dagger}}
\newcommand{\C}[0]{\mathcal{C}}

\newcommand{\rhot}[0]{\hat{\rho}(t)}
\renewcommand{\a}[0]{\hat{a}}
\renewcommand{\H}[0]{\hat{H}}
\newcommand{\sss}[0]{\hat{\rho}_{\rm ss}}

\newsavebox{\mstrut}
\newcommand{\bbra}[1]{
    \sbox{\mstrut}{\(#1\)}
    \mathinner{\left\langle\kern-0.3\ht\mstrut\left\langle{#1}\right|\mkern-2mu\right|}
}
\newcommand{\kett}[1]{
    \sbox{\mstrut}{\(#1\)}
    \mathinner{\left|\mkern-2mu\left|{#1}\right\rangle\kern-0.3\ht\mstrut\right\rangle}
}

\newcommand{\kettbbra}[2]{
    \sbox{\mstrut}{\(#1\)}
    \mathinner{\left|\mkern-2mu\left|{#1}\right\rangle\kern-0.3\ht\mstrut\right\rangle\mkern-5.8mu\left\langle\kern-0.3\ht\mstrut\left\langle{#2}\right|\mkern-2mu\right|}
}

\usepackage[normalem]{ulem}
\renewcommand\sout{\bgroup\markoverwith{\textcolor{red}{\rule[0.5ex]{2pt}{0.8pt}}}\ULon}

\begin{document}

\author{Luca Gravina}
\affiliation{Institute of Physics, Ecole Polytechnique Fédérale de Lausanne (EPFL), CH-1015 Lausanne, Switzerland}
\affiliation{Center for Quantum Science and Engineering, Ecole Polytechnique Fédérale de Lausanne (EPFL), CH-1015 Lausanne, Switzerland}
\author{Fabrizio Minganti}
\affiliation{Institute of Physics, Ecole Polytechnique Fédérale de Lausanne (EPFL), CH-1015 Lausanne, Switzerland}
\affiliation{Center for Quantum Science and Engineering, Ecole Polytechnique Fédérale de Lausanne (EPFL), CH-1015 Lausanne, Switzerland}
\author{Vincenzo Savona}
\affiliation{Institute of Physics, Ecole Polytechnique Fédérale de Lausanne (EPFL), CH-1015 Lausanne, Switzerland}
\affiliation{Center for Quantum Science and Engineering, Ecole Polytechnique Fédérale de Lausanne (EPFL), CH-1015 Lausanne, Switzerland}

\title{A critical Schrödinger cat qubit}

\begin{abstract}
Encoding quantum information onto bosonic systems is a promising route to quantum error correction. 
In a cat code, this encoding relies on the confinement of the system's dynamics onto the two-dimensional manifold spanned by Schrödinger cats of opposite parity.
In \textit{dissipative cat qubits}, an engineered dissipation scheme combining two-photon drive and two-photon loss has been used to autonomously stabilize this manifold, ensuring passive protection against, e.g., bit-flip errors regardless of their origin.
Similarly, in \textit{Kerr cat qubits}, where highly-performing gates can be engineered, two-photon drive and Kerr nonlinearity cooperate to confine the system to a two-fold degenerate ground state manifold spanned by cat states of opposite parity. 
Dissipative, Hamiltonian, and hybrid confinement mechanisms have been investigated at resonance, i.e., for driving frequencies matching that of the cavity.
Here, we propose a \textit{critical cat code}, where both two-photon loss and Kerr nonlinearity are present, and the two-photon drive is allowed to be out of resonance.
The performance of this code is assessed via the spectral theory of Liouvillians in all configurations ranging from the purely dissipative to the Kerr limit.
We show that large detunings and small, but non-negligible, two-photon loss rates are fundamental to achieve optimal performance. We further demonstrate that the competition between nonlinearity and detuning results in a first-order dissipative phase transition, leading to a squeezed vacuum steady state.
We show that to achieve the maximal suppression of the logical bit-flip rate requires initializing the system in the metastable state emerging from the first-order transition, and we detail a protocol to do so.
Efficiently operating over a broad range of detuning values, the critical cat code is particularly resistant to random frequency shifts characterizing multiple-qubit operations, opening venues for the realization of reliable protocols for scalable and concatenated bosonic qubit architectures.
\end{abstract}

\date{\today}

\maketitle

\section{Introduction}
\label{sec:intro}

The development of large-scale quantum computers relies on the possibility of taming errors, i.e., the irreversible processes stemming from the interaction of the system with its surrounding environment \cite{lidar2013, haroche2013, preskill2018, breuer2007}.
Quantum error correction schemes redundantly encode quantum information onto multi-level quantum systems, in a way that enables to detect and correct specific types of errors without affecting the stored quantum information \cite{nielsen2011,campbell2017,terhal2015}.
The mainstream quantum error correction paradigm consists in encoding the $\ket{0_L}$ and $\ket{1_L}$ logical states onto a two-dimensional subspace of the Hilbert space characterizing the states of several physical qubits \cite{nielsen2011, lidar2013}. 
Despite its promise of scalability, this type of encoding suffers from its large hardware footprint and the high connectivity between physical qubits required to execute fault-tolerant quantum computations.

An alternative paradigm to detect and correct quantum errors consists in encoding the logical states of a qubit onto an appropriately selected subspace of the Hilbert space of a harmonic oscillator \cite{gottesman2001, mirrahimi2014, cai2021, joshi2021, terhal2020, knill2000, michael2016}. These \textit{bosonic quantum codes} are characterized by a reduced hardware footprint, and essentially eliminate the daunting challenge of simultaneously controlling the multiple degrees of freedom of several physical qubits \cite{albert2018}.

In these systems, information is encoded as a symmetric pattern in phase space \cite{joshi2021}. While a translational symmetry underlies the GKP code \cite{gottesman2001}, a rotational one characterizes the \emph{Schr\"odinger cat code}, where the logical manifold is spanned by cat states of opposite parity \cite{gilles1994, mirrahimi2014}.

Schr\"odinger cat qubits have been realized, in particular, on superconducting circuit platforms which are mainly prone to two noise processes: single particle loss and dephasing \cite{joshi2021}.
These directly map onto errors on the logical qubit, namely, bit- and phase-flip errors.
Confinement of the system's dynamics to the cat manifold relies on a subtle interplay between engineered parametric processes. In recent years two approaches have been proposed to achieve this confinement.

\emph{Dissipative confinement} relies on an engineered dissipation scheme combining two-photon drive $G$ and two-photon loss $\eta$ \cite{mirrahimi2014, leghtas2015, touzard2018, xu2022, goto2016, cochrane1999} to generate and autonomously stabilize the code manifold.
Dissipative cats are intrinsically resistant to leakage processes, ensuring an exponential suppression of the phase-flip error rate in the cat's photon number \cite{mirrahimi2014, leghtas2015, albert2016, albert2018a}. 
Their main drawback is the limited performance of logical gates on current superconducting platforms \cite{gautier2022, touzard2018}.

\emph{Hamiltonian confinement}, on the other hand, relies on the Kerr nonlinearity $U$ to restrict the system to the doubly degenerate ground space of the Kerr parametric oscillator \cite{puri2017,puri2019, puri2020, grimm2020}.
Gate performance can be improved by the application of, e.g., super-adiabatic pulse designs \cite{xu2022}, limiting the amount of leakage induced by gate operations. Nevertheless, this protocol remains susceptible to thermal and dephasing noise, which are no longer exponentially suppressed in the photon number \cite{gautier2022}. This issue is to be ascribed to the the level structure of the Kerr parametric oscillator, and  has been identified as the primary limitation of a
\textit{hybrid} confinement scheme, i.e one combining sizable two-photon loss and Kerr-nonlinearity. A new confinement scheme  inheriting its nonlinearity from a two-photon exchange with an external two-level system \cite{gautier2022} was recently proposed as a possible solution to this issue.
Note that dissipative, Hamiltonian, and the latter proposal alike all operate in a \textit{resonant} regime, i.e., one where the pump-to-cavity detuning is $\Delta=0$. Only very recently, finite values of $\Delta$ have been explored, albeit restricting to the limiting case of Hamiltonian confinement, in Refs.~\cite{ruiz2022, frattini2022, Venkatraman2022}.

Here, we adopt a novel perspective, operating the cat in a hybrid regime and in the presence of sizable detuning. We dub our encoding a \textit{critical cat code} by virtue of the first-order dissipative phase transition (DPT) characterizing the system. 

By exploring the parameter space, we prove that
\begin{itemize}
    \item Efficient quantum information encoding is not limited to $\Delta=0$ if $U \neq 0$, but extends over a broad range of values of $\Delta$, where the qubit can be operated (Sec.~\ref{sec:errors}).
    \item The critical cat \textit{outperforms} its Hamiltonian, dissipative, and hybrid-resonant counterparts as large photon numbers and an \textit{enhanced exponential suppression} of bit-flip errors can be achieved for carefully chosen  values of $\Delta$ (Sec.~\ref{sec:errors}).
    \item  The critical cat qubit provides a significant step forward in the hybrid operation of cat qubits, overcoming many of the limitations of the protocol detailed in Ref~\cite{gautier2022}, such as the need for additional confinement Hamiltonians other than the Kerr parametric oscillator (with annexed hardware overhead) and lack of spontaneous stabilization (Sec.~\ref{sec:errors}).
    \item 
    The code may achieve optimal performance in a regime where the code space is only metastable, while the steady state is a squeezed vacuum.     
    In this regime, emerging as a consequence of the aforementioned DPT, logical errors compete with the leakage process characterizing the decay from the code space to the vacuum. 
    Acknowledging the presence of criticality is thus fundamental for the proper characterization and initialization of detuned cat qubits, regardless of their regime of operation (Sec.~\ref{sec:encoding}).  
    \item Parameter configurations exist for which the qubit becomes resilient against uncontrolled changes in its frequency (Sec.~\ref{sec:Frequency_shift}), originating, for instance, from the dispersive coupling to external reservoirs or ancillary circuital elements \cite{lieu2020, chamberland2022}. The latter being typically required for the realization of two-qubit entangling gates or concatenated-qubit error-correction protocols.  
\end{itemize}

The article is structured as follows. 
In Sec.~\ref{sec:cat_gen} we introduce the cat qubit, the dissipative and Hamiltonian processes underlying its generation, and we briefly review the role of Liouvillian symmetries in bosonic quantum information encoding.
In Sec.~\ref{sec:errors} we analyze the effect of detuning on a hybrid cat code, demonstrating the critical cat's enhanced resilience to bit-flip errors.
In Sec.~\ref{sec:encoding} we show the existence of metastable timescales and propose a protocol for the efficient initialization of detuned cats.
We study the resilience to frequency-shift errors in Sec.~\ref{sec:Frequency_shift} and discuss protocols for bias-preserving gates in Sec.~\ref{sec:gates}. We draw our conclusions and discuss future perspectives in Sec.~\ref{Sec:Conclusions}.
Appendix~\ref{app:encoding} presents details on the calculations underlying the different encoding mechanisms detailed in Sec.~\ref{sec:QIencoding}.

\section{Generation and stabilization of Schr\"odinger cats}
\label{sec:cat_gen}
Schr\"odinger cat states are the even and odd superpositions of two coherent states of opposite phases
\begin{equation}
\label{Eq:definition_cat}
\ket{\C_\alpha^\pm} = \frac{\ket{\alpha} \pm \ket{-\alpha}  }{2\sqrt{1 \pm e^{-2|\alpha|^2}}},
\end{equation}
where the coherent states $\ket{\alpha}$ are eigenstates of the annihilation operator $\a$, i.e. $\a \ket{\alpha} = \alpha \ket{\alpha}$ \cite{gerry2004, haroche2013}.
When expressed in the number (Fock) basis, $\ket{\C_\alpha^\pm}$ contain only components with respectively even and odd number of photons. As such, they are eigenstates of the parity operator $\hat{\Pi} = \exp(i \pi \a^\dagger \a)$ with eigenvalues $\pm 1$. 

Cat qubits are encoded in the two-dimensional manifold defined by cat states of opposite parity. The prevailing convention defining the logical basis as \footnote{Note that this choice is arbitrary. An equivalently popular convention defines the logical basis as $ \ket{0_L} =  \ket{\C_\alpha^+}\,,\quad \ket{1_L} =  \ket{\C_\alpha^-} $. This amounts to a $\pi/2$ rotation within the logical space and an exchange between bit and phase-flip errors.
}
\begin{equation}
    \begin{aligned}
        \ket{0_L} &= \frac{1}{\sqrt{2}}\qty(\ket{\C_\alpha^+} + \ket{\C_\alpha^-}) \approx \ket{+\alpha} + \order{e^{-2|\alpha|^2}}\\
        \ket{1_L} &= \frac{1}{\sqrt{2}}\qty(\ket{\C_\alpha^+} - \ket{\C_\alpha^-}) \approx \ket{-\alpha} + \order{e^{-2|\alpha|^2}},
    \end{aligned}
\end{equation}
so that $\ket{\pm_L} = \ket{\C_\alpha^\pm}$. The generation and stabilization of Schrödinger cat codes heavily relies on parity-preserving processes involving exclusively the pairwise exchange of photons between the system and its environment \cite{mirrahimi2014, gilles1994, hachiii1994}.
\begin{figure}[bt]
\includegraphics[width=0.5\textwidth]{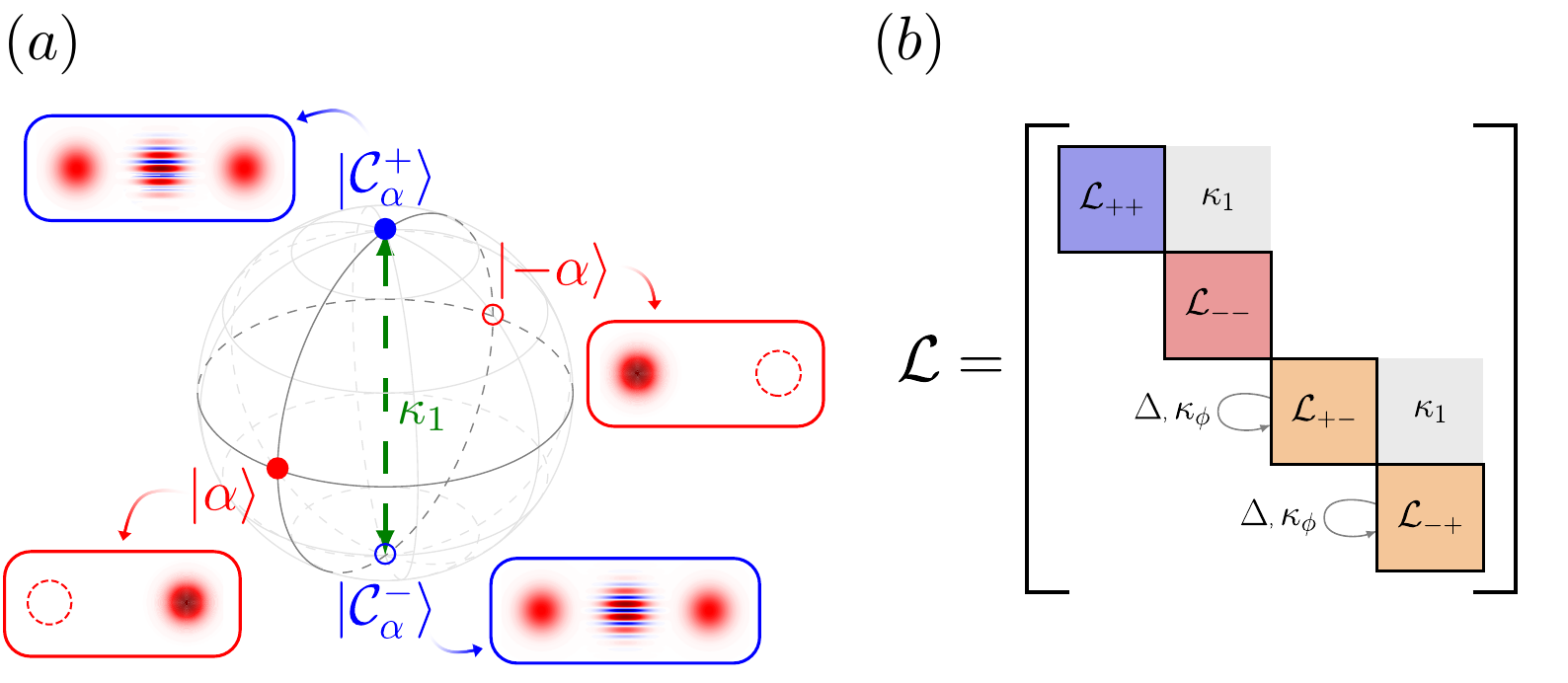} 
\caption{\label{fig:Catqubit} (a) The even and odd cat states, represented by their Wigner functions, encode the logical states $\ket{0_L}$ and $\ket{1_L}$. Their superpositions, approximately the two coherent states $\ket{\pm \alpha}$, define the logical states $\ket{\pm_L}$. 
(b) Block structure of the Liouvillian arising from its strong $\mathcal{Z}_2$ symmetry. 
Finite values of $\Delta$ or $\kappa_\phi$ act only within the blocks $\LL_0^{+-}$ and $\LL_0^{-+}$, so that  the four-block-diagonal structure of $\LL_0$ is preserved. The single-photon loss $\kappa_1$, instead, connects different blocks, resulting in a two-block-diagonal structure typical of weakly $\mathcal{Z}_2$-symmetric systems.} 
\end{figure}
These processes are modeled by the effective Lindblad master equation \cite{breuer2007, gorini1976, gorini1978, lindblad1976}:
\begin{subequations}
\label{eq:full_cat}
\begin{align}
\allowdisplaybreaks
&\partial_t \hat{\rho}= \LL \hat{\rho} = \LL_0 \hat{\rho} + \LL_1(\kappa_1,\kappa_\phi) \hat{\rho}\\[0.15cm]
&\LL_0 \hat{\rho} = -i\comm{\H}{\hat{\rho}} + \eta\D\left[\a^2\right] \hat{\rho} \label{eq:full_cat_subeq2}\\[0.15cm]
&\H=\Delta\,\a^\dagger \a +  \frac{G}{2} \qty( \a^{2} + \a^{\dagger2})-\frac{U}{2} \a^{\dagger2} \a^{2}\\[0.15cm]
&\LL_1(\kappa_1,\kappa_\phi) \hat{\rho} = \kappa_1 \D[\a]\, \hat{\rho}+ \kappa_\phi\D[\a^\dagger \a]\, \hat{\rho}.
\end{align}
\end{subequations}
Here, $\hat{\rho} \equiv \rhot$ (for brevity) is the system density matrix at time $t$, and $\LL$ is the Liouvillian superoperator which we separate into two parts: $\LL_0$ and $\LL_1(\kappa_1,\kappa_\phi)$. The former, $\LL_0$, confines the dynamics to the cat manifold. Its Hamiltonian $\H$ accounts for the two-photon driving field amplitude $G$, the Kerr nonlinearity $U$, and the pump-to-cavity detuning $\Delta$. $\LL_0$ also accounts for two-photon losses, described by the rate $\eta$ at which pairs of photons are incoherently emitted from the cavity.
The latter, $\LL_1(\kappa_1,\kappa_\phi)$, describes the unwanted effects of the environment on the code, inducing errors in the logical qubit. Here, $\kappa_1$ represents the single-photon loss rate, while $\kappa_\phi$ is the total dephasing rate.

The dissipator $\D[\hat{J}]$ describes the action of the jump operator $\hat J$ on $\hat\rho$, and is defined as
\begin{equation}
    D[\hat{J}] \hat{\rho} = \hat{J}\hat{\rho} \hat{J}^\dagger - \frac{\hat{J}^\dagger \hat{J} \hat{\rho} + \hat{\rho} \hat{J}^\dagger \hat{J}}{2}.
\end{equation}
The combined effect of $\LL_0$ and $\LL_1$ is usually recast in terms of a phase-flip error rate $\Gamma_\phi$, and a bit-flip error rate $\Gamma$ acting on the logical qubit \cite{mirrahimi2014}. It is known \cite{lescanne2020, guillaud2019, guillaud2021, mirrahimi2014, gautier2022, ruiz2022, chamberland2022} that the main advantage of cat qubits lies in the exponential suppression of $\Gamma$ in the average photon number, with $\Gamma_\phi$ increasing only linearly. 
This is what drives the interest in cat states as biased-noise codes.
As we will show below, however, this picture is incomplete and, in most detuned regimes of operation, at least one additional rate $\Gamma_{\rm leak}$ must be introduced, characterizing the leakage towards states outside the logical qubit manifold.

\subsection{Symmetries}

Liouvillian symmetries are distinguished into two classes: \emph{weak} and \emph{strong} \cite{blume-kohout2010, buca2012, lieu2020, sanchezmunoz2018, sanchezmunoz2019, thingna2021, zhang2020}.
A weak symmetry occurs when an operator $\hat{O}$ can be found that obeys 
\begin{equation}
    \LL(\hat{O} \hat{\rho}\, \hat{O}^\dagger) = \hat{O}(\LL \hat{\rho} )\hat{O}^\dagger,
\end{equation} with $\LL\hat\rho = -i\comm*{\H}{\hat\rho} + \sum_i\gamma_i\mathcal{D}[\hat J_i]\hat\rho$.
A strong symmetry, on the other hand, requires 
\begin{equation}
    \comm{\hat{O}}{\H}=\comm{\hat{O}}{\hat{J}_i}=0.
\end{equation}
While a weak symmetry only guarantees the existence of a single steady state, a strong symmetry induces an $n$-dimensional steady-state manifold, with $n$ the number of non-equivalent irreducible representations of the symmetry group \cite{zhang2020}.

The ideal Liouvillian $\LL_0$ described in Eq.~\eqref{eq:full_cat_subeq2} is characterized by a \textit{strong} $\mathcal{Z}_2$ symmetry, as both the Hamiltonian and the jump operators commute with the parity operator $\hat{\Pi}$.
The Liouvillian can thus be written in the block diagonal form $\LL_0 = \LL_0^{++} \oplus \LL_0^{+-} \oplus \LL_0^{-+} \oplus \LL_0^{--}$  sketched in Fig.~\ref{fig:Catqubit}(b) \cite{albert2014,buca2019}. 
The dynamics within each block is conveniently described in terms of its eigenvalues $\lambda^{\mu\nu}_{j}$ and right eigenoperators $\hat{\rho}^{\mu\nu}_j$ where
\begin{equation}
\label{Eq:block_eigenvalues}
 \LL_0^{\mu\nu} \hat{\rho}^{\mu\nu}_j = \lambda^{\mu\nu}_{j}  \hat{\rho}^{\mu\nu}_j \quad\text{with}\,\,-\Re{\lambda^{\mu\nu}_{j}}<-\Re{\lambda^{\mu\nu}_{j+1}}.
\end{equation}
Due to the strong symmetry, the latter are also eigenoperators of $\hat \Pi$ according to the relation
\begin{equation}
    \label{eq:coherences}
    \hat{\Pi} \hat{\rho}^{\mu\nu}_{j} = \mu \hat{\rho}^{\mu\nu}_{j}, \quad {\rm and } \quad   \hat{\rho}^{\mu\nu}_{j} \hat{\Pi}^\dagger= \nu \hat{\rho}^{\mu\nu}_{j}.
\end{equation}
From here on, we shall refer to $\hat{\rho}_0^{\mu \mu}$ as \textit{steady states} and to $\hat{\rho}_0^{\mu \nu}$, with $\mu\neq\nu$, as \textit{coherences} \cite{albert2014, albert2018a}.
While the imaginary part of $\lambda^{\mu\nu}_{j}$ describes the oscillations within the specified sector, its negative real part sets the relaxation rate towards the steady state. Throughout the paper, we deem
\begin{equation}
    \label{eq:decay_rate}
    \Lambda^{\mu \nu}_{j} = -\Re{\lambda^{\mu\nu}_{j}}.
\end{equation}

Complementary information on the dynamical properties of $\LL$ can be retrieved from its left eigenoperators $\hat{\sigma}^{\mu\nu}_j$, defined as
\begin{equation}
\LL_{\mu\nu}^\dagger \hat{\sigma}^{\mu\nu}_j = \left(\lambda^{\mu\nu}_{j}\right)^*  \hat{\sigma}^{\mu\nu}_j.
\end{equation}
They are related to $\hat{\rho}^{\mu\nu}_j$ by the bi-orthogonality relation
\begin{equation}
    \Tr{{(\hat{\sigma}^{\mu\nu}_k})^\dagger  \hat{\rho}^{\mu^{\prime}\nu^{\prime}}_j} = \delta_{\mu, \mu^{\prime}} \delta_{\nu, \nu^{\prime}} \delta_{j,k},
\end{equation}
and form a basis to express the time-evolution of observables in the Heisenberg picture. The null-eigenoperators $\{\hat{\sigma}^{\mu\nu}_0\}_{\mu,\nu}$ of $\LL^\dagger$, in particular, are conserved quantities, defining the set of all observables which remain constant throughout the evolution \cite{albert2014}.

Upon introducing single-particle loss errors with rate $\kappa_1$, the strong $\mathcal{Z}_2$ symmetry is lost as the four-block-diagonal structure of the Liouvillian is replaced by the two-block structure $\LL = \LL_+ \oplus \LL_-$ typical of the \textit{weak} $\mathcal{Z}_2$ symmetry that persists in the system [see Fig.~\ref{fig:Catqubit}(b)].
We denote the eigenvalues of the weakly $\mathcal{Z}_2$ symmetric Liouvillian as
\begin{equation} 
\label{Eq:block_eigenvalues_weak}
 \LL_{\mu} \hat{\rho}^{\mu}_j = \lambda^{\mu}_{j}  \hat{\rho}^{\mu}_j \quad\text{with}\,\,-\Re{\lambda^{\mu}_{j}}<-\Re{\lambda^{\mu}_{j+1}}.
\end{equation}
Similarly, the left eigenoperators are defined as 
\begin{equation}
\LL_{\mu}^\dagger \hat{\sigma}^{\mu}_j = (\lambda^{\mu}_{j})^*  \hat{\sigma}^{\mu}_j, 
\end{equation}
and in analogy with Eq.~\eqref{eq:decay_rate} we define the rates
\begin{equation}
    \label{eq:decay_rate_weak}
    \Lambda^{\mu}_{j} = -\Re{\lambda^{\mu}_{j}}.
\end{equation}
Within this picture, an arbitrary state \textit{will decay towards the system's unique steady state}.

\subsection{Quantum information encoding}
\label{sec:QIencoding}

An open quantum system can encode quantum information and perform quantum computation 
(details in Appendix~\ref{app:encoding}) if, at time $t=0$,
the following relation holds
\begin{equation}\label{Eq:NS_main}
    \hat{\rho}(0) \equiv \hat{Q}(0) \otimes \hat{M}(0).
\end{equation}
The matrix $\hat{\rho}(0)$ is that of a bipartite system, where $\hat{Q}(0)$ is the state encoding quantum information, while $\hat{M}(0)$ is a mixed state. 
Hamiltonian and dissipative processes will make the system evolve into 
\begin{equation}\label{Eq:general_time_evolution}
    \hat{\rho}(t) = \beta(t) \hat{Q}(t) \otimes \hat{M}(t) + [1-\beta(t)]\hat{\rho}_{\rm leak}(t),
\end{equation}
where $\hat{\rho}_{\rm leak}$ represents the leakage outside the code space, $\beta(t)$ is determined by the Liouvillian dynamics, and we are neglecting possible coherences between the code and the leakage space.
If $\beta(t)=1$, the Liouvillian dynamics will not induce leakage.
In this case, the evolution $Q(t)$ represents that of a logical qubit in the presence of logical bit- ($\Gamma$) and phase-flip ($\Gamma_{\phi}$) errors.
The special case of $\hat{Q}(0) = \hat{Q}(t) $, i.e.,
\begin{equation}\label{Eq:noiseless_encoding}
    \hat{\rho}(t) = \hat{Q}(0) \otimes \hat{M}(t),
\end{equation}
corresponds to $\Gamma = \Gamma_{\phi}=0$.
This is a \textit{noiseless subsystem} \cite{lidar2013, lidar2014, albert2016, buca2019, kempe2001, knill2000, lidar1998}, and can only emerge  in the system under investigation if $\kappa_1 = \kappa_\phi=\Delta =0$ \cite{mirrahimi2014, lieu2020}.
Indeed, only when a zero Liouvillian eigenvalue exists within each  symmetry sector, i.e., $\Lambda^{\mu \nu}_{0} =0$ $\forall\mu,\nu\in\{\pm\}$, quantum information can be stored indefinitely, as this guarantees infinitely long lived steady states and coherences.
If, instead, $\kappa_1, \, \kappa_\phi, \, \Delta \neq 0$, and $\beta(t)=1$, the system will evolve towards its unique steady state as
\begin{equation}\label{Eq:statedy-state_encoding}
    \hat{\rho}(t) = \hat{Q}(t) \otimes \hat{M}(t).
\end{equation}
For the logical qubit, this corresponds to finite values of $\Gamma$ and $\Gamma_\phi$ and will result in the irreversible evolution of the qubit towards a steady state at the center of the Bloch sphere.
We will refer to this case as the \textit{steady state encoding}.

If, instead, $\beta(t)<1$, the system will leak out of the code space.
This case implies the existence of at least one additional timescale $1/\Gamma_{\rm leak}$, governing the variation of $\beta(t)$.
A limiting case in this scenario, which we will call a \textit{metastable encoding}, is
\begin{equation}\label{Eq:metastable_encoding}
\begin{split}
    \rhot =& \beta(t) \hat{Q}(t) \otimes \hat{M}(t) + [1-\beta(t)]\sss, \\
    \beta(t) =& 1-e^{\Gamma_{\rm leak} t}.
\end{split}    
\end{equation}
As we will show below, an optimal biased-noise metastable encoding -- wherein quantum information can be efficiently stored -- 
requires $\Gamma$ and $\Gamma_{\rm leak}$ to be very small with respect to the typical rate of gate operations.

\section{Enhanced suppression of bit-flip errors}
\label{sec:errors}

Dissipative, Hamiltonian, and hybrid stabilization alike, have mainly been investigated at $\Delta = 0$, where, in the absence of errors, Eq.~\eqref{Eq:NS_main} is exactly met. 
In this section we unveil $\Delta$ as a new control parameter, demonstrating its potential to significantly enhance the performance of cat codes. 

As we are interested in the cat code as a biased-noise encoding,
we focus in this section on finding the optimal parameter configuration minimizing the bit-flip error rate $\Gamma$ in the presence of both photon-loss and dephasing errors. 
For consistency with the existing literature \cite{ruiz2022, gautier2022, grimm2020, chamberland2022, frattini2022}, we  take $0< G < 8$ and consider throughout this section two noise configurations: one dominated by photon-loss ($\kappa_1 = 10^{-3}\gg \kappa_\phi= 10^{-5}$), the other by dephasing ($\kappa_1 = 10^{-5}\ll \kappa_\phi= 10^{-3}$).
We defer a thorough analysis of the interplay between $\Gamma$ and $\Gamma_{\rm leak}$, and of the different encoding emerging at finite detuning, to Sec.~\ref{sec:encoding} and Appendix~\ref{app:encoding}.

\subsection{Liouvillian analysis of critical cat}
\label{sec:spectral}

\begin{figure}[htb]
\includegraphics[width=0.48\textwidth]{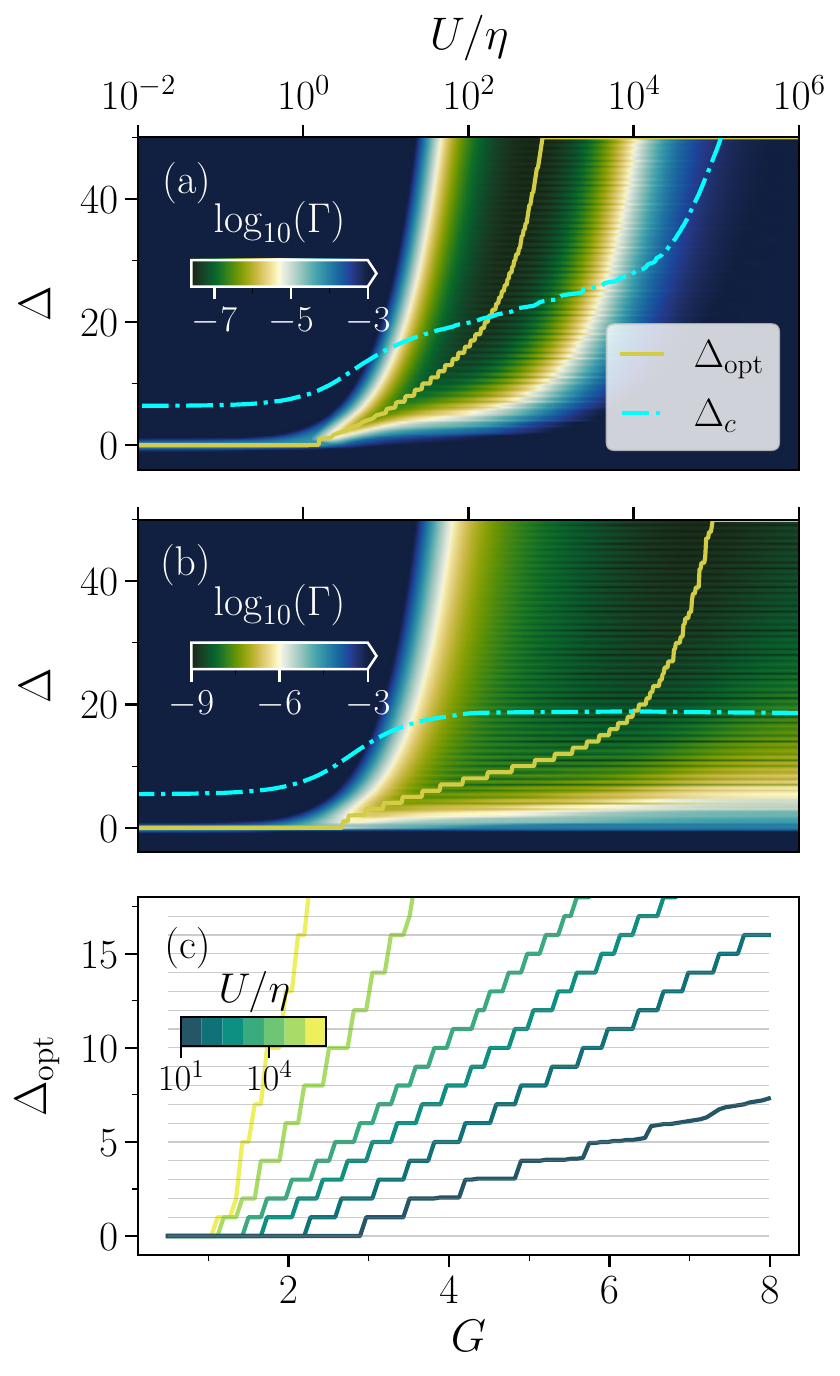} 
\caption{\label{fig:gamma_v_Delta_v_U} 
Bit-flip error rate $\Gamma$ as a function of the detuning $\Delta$ and the ratio $U/\eta=\tan(\theta)$ defined in Eq.~\eqref{Eq:W_definition}.  The dissipative limit corresponds to $\theta=0$ ($U=0$, $\eta=1$). The opposite limit, i.e the Kerr (or Hamiltonian) limit, is reached for $\theta=\pi/2$ ($U=1$, $\eta=0$). The optimal detuning values $\Delta_{\rm opt}(G=5,\theta)$ are shown as continuous yellow lines. The dot-dashed line denotes the critical threshold $\Delta_{c}$ separating the phase with large steady-state photon number from the vacuum one (see Sec.~\ref{sec:encoding}). We set the two-photon driving field amplitude $G=5$, and $\kappa_1 = 10^{-5}\ll \kappa_\phi= 10^{-3}$ (a) and $\kappa_1 = 10^{-3}\gg \kappa_\phi= 10^{-5}$ (b). 
(c) $\Delta_{\rm opt}(G,\theta)$ as a function of $G$ for different values of $\theta$ and $\kappa_1 = 10^{-3}\gg \kappa_\phi= 10^{-5}$.}
\end{figure}

In this section, we analyze the spectral properties of $\LL = \LL_0 + \LL_1$ [Eq.~\eqref{eq:full_cat}] to gain insight into the interplay between $\Delta$, $U$ and/or $\eta$ in relation to $\Gamma$. We relate these parameters by setting
\begin{equation} 
\label{Eq:W_definition}
    \eta = W \cos(\theta), \quad  U = W \sin(\theta), \quad \eta^2 + U^2 = W^2,
\end{equation}
where $W$ represents the natural unit of the system. In what follows all energies and times will be reported in units of $W$ and $W^{-1}$ respectively. By varying $\theta$ between $0$ and $\pi/2$, we are able to continuously explore the hybrid region between the dissipative and Hamiltonian (Kerr) limits. 
We do so in Fig.~\ref{fig:gamma_v_Delta_v_U} where, for a fixed value of $G$, we explore the phase diagram of $\Lambda_0^-$ in $\Delta$ and $U/\eta=\tan(\theta)$ for both noise configurations.
Within this Liouvillian picture, $\Gamma \equiv\Lambda_0^{-}$ represents the slowest timescale involved in the decay of the field quadratures. For instance, it describes the rate at which $\ket{\alpha} \to \ket{-\alpha}$ whenever $\Delta=0$ \cite{frattini2022, albert2018a}.

While in the dissipative limit ($U/\eta<1$) the detuning configuration minimizing $\Gamma$ is the resonant one ($\Delta_{\rm opt}=0$), for sizable Kerr nonlinearities ($U/\eta>1$) $\Delta_{\rm opt}>0$. Indeed, in this regime, the introduction of a nonvanishing detuning significantly suppresses the value of $\Gamma$ not only with respect to the physical error rates, but to all resonant configurations as well. This result is at the heart of the \emph{critical cat code}, as it highlights the key role of detuning in suppressing bit-flip errors, with its effectiveness being linked to the presence of a non-zero Kerr nonlinearity (see Fig.~\ref{fig:gamma_v_Delta_v_U}).

\begin{figure}[tb]
\center
\includegraphics[width=0.49\textwidth]{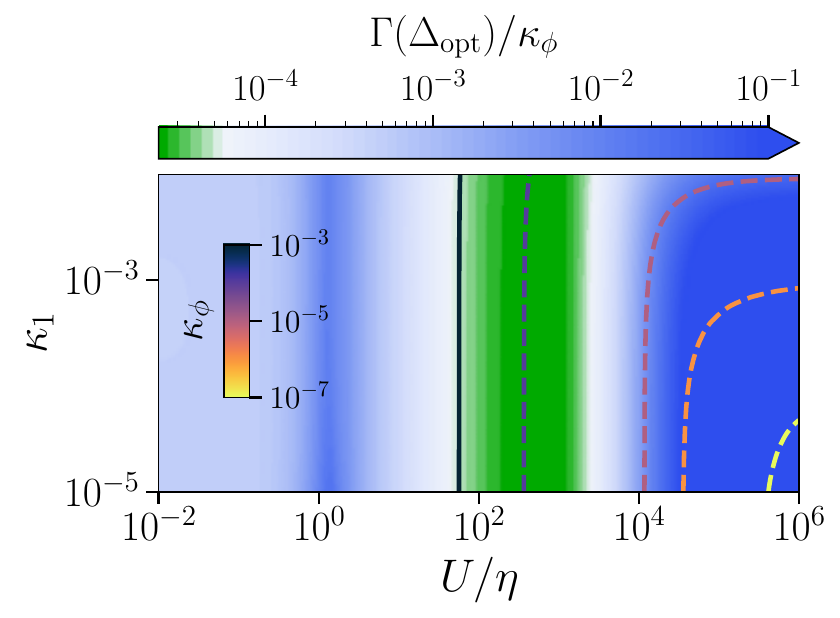}
\caption{\label{fig:deltaopt_v_k1_v_U} Optimal bit-flip error rate $\Gamma(\Delta_{\rm opt})$ as a function of the single-photon loss rate $\kappa_1$ and of the ratio $U/\eta$ for $\kappa_\phi=10^{-3}$. The value of $\Gamma(\Delta_{opt})$ is normalized to $\kappa_\phi$ to ensure consistency in the limiting values of the displayed quantity for all choices of $\kappa_\phi$. The solid line highlights the lowest-$(U/\eta)$ isoline $\Gamma(\Delta_{\rm opt})/\kappa_\phi = 5\cdot10^{-5}$. The dashed curves display the same quantity for  $\kappa_\phi = 10^{-4}, 10^{-5}, 10^{-6}, 10^{-7}$. The optimal configuration is clearly seen to shift towards the Kerr limit for increasing values of $\kappa_\phi$. }
\end{figure}

In this direction, comparison of Figs.~\ref{fig:gamma_v_Delta_v_U}(a) and (b) shows that, in a configuration dominated by dephasing, the optimal $\Gamma$ is found deep in the hybrid regime ($U/\eta \approx 10^2$), while in a configuration dominated by photon loss it is achieved closer to the Kerr limit ($U/\eta \gtrsim 10^5$). 
This is further supported by Fig.~\ref{fig:deltaopt_v_k1_v_U}, which displays $\Gamma(\Delta_{\rm opt})$ as a function of $\kappa_1$ and $U/\eta$, for different values of $\kappa_\phi$. The data reveals that the region with optimal protection from errors progressively shifts from the hybrid-phase to the Kerr limit as $\kappa_\phi$ is decreased. This trend can be observed for all values of $\kappa_1$, though it becomes more pronounced for $\kappa_1\gg\kappa_\phi$.
Note that although  $\Delta_{\rm opt}$ is rapidly increasing in $U/\eta$ (yellow solid lines), for both noise configurations we find large regions in the parameter space where $\Gamma$ does not significantly depart from its optimal value. While in the photon-loss dominated case this region extends from a hybrid to a purely-Kerr regime [Fig.~\ref{fig:gamma_v_Delta_v_U}(b)], in the dephasing-dominated one it always requires nonvanishing $\eta$ [Fig.~\ref{fig:gamma_v_Delta_v_U}(a)]. Overall this introduces a tradeoff between the choice of $\theta$ and the attainable values of $\Delta$ which is an experimentally finite resource. Indeed, for very large detuning, the approximations leading from a microscopic description to the effective model in Eq.~\eqref{eq:full_cat} break down.
For this reason, we restrict our considerations to $\Delta\leq\Delta_{\rm max}=50$, consistently with the most recent experimental efforts in this direction \cite{Venkatraman2022}.
In Fig.~\ref{fig:gamma_v_Delta_v_U}(c) we characterize the behavior of $\Delta_{\rm opt}$ as a function of $G$ for various values of $U/\eta$. The increase of $\Delta_{\rm opt}$ follows a staircase pattern taking on integer multiples of $U$ for which additional degeneracies in the spectrum of the Kerr parametric oscillator have been found \cite{frattini2022,Venkatraman2022,ruiz2022}. This staircase pattern is particularly pronounced in the hybrid regime, becoming steeper in the Kerr limit, consistently with the aforementioned divergence of $\Delta_{\rm opt}$ in $U/\eta$.

 \begin{figure}[tb]
\includegraphics[width=0.48\textwidth]{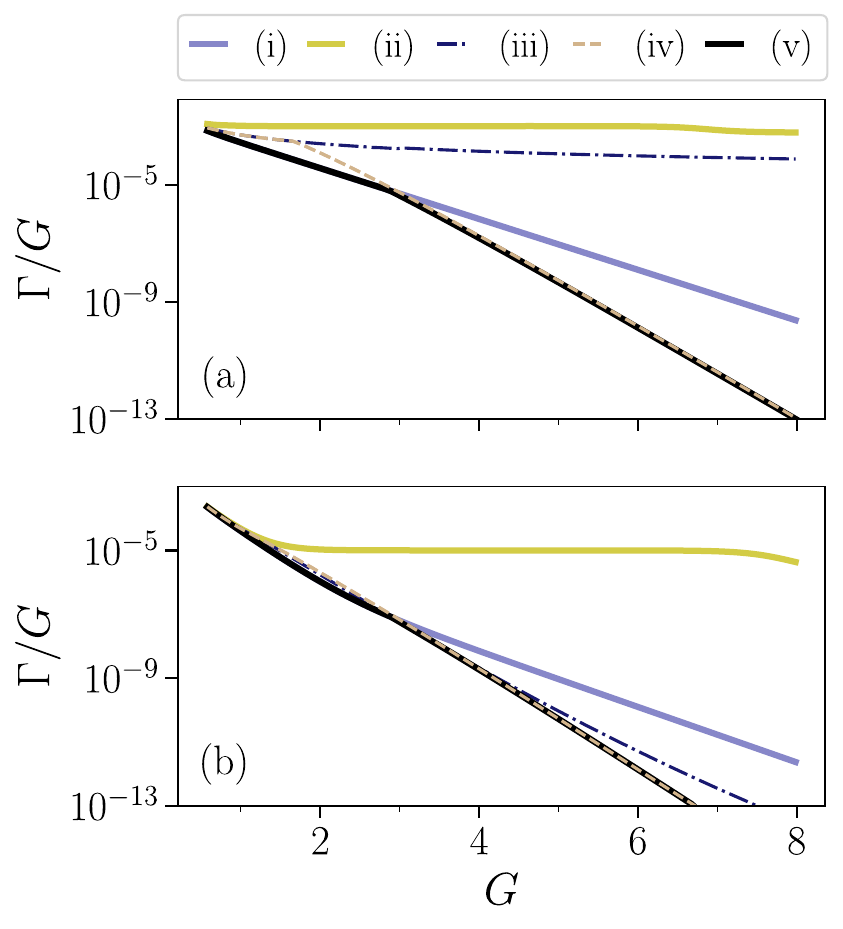} 
\caption{\label{fig:gamma_v_G} 
Bit-flip error rate $\Gamma$ as a function of the driving field amplitude $G$ for different values of $\Delta$ and $\theta$, for $\kappa_1 = 10^{-5}\ll \kappa_\phi= 10^{-3}$ (a), $\kappa_1 = 10^{-3}\gg \kappa_\phi= 10^{-5}$ (b). In both panels, we display results for (i) the Kerr limit $U/\eta=10^{6}$, (ii) the dissipative limit $U/\eta=0$, (iii) the optimal configuration obtained by optimizing both $\theta$ and $\Delta$, (iv) the result obtained by setting $U/\eta=10^{6}$ and optimizing over $\Delta$. The curve denoted as (v) was obtained by setting $U/\eta=10^2$ (a) and $U/\eta=10^5$ in (b).}
\end{figure}

\subsection{Scaling of bit-flip error rates}
Having shown the advantage of the hybrid critical encoding for $G=5$, we set out to demonstrate the enhanced exponential suppression of $\Gamma$ with the average photon number $\expval*{\ad\a}\propto G$.
In Fig.~\ref{fig:gamma_v_G}, we assess the scaling of $\Gamma$ as a function of $G$ for several values of $\theta$ and $\Delta$.  As baselines for the assessment of the critical cat, we consider the resonant dissipative (blue) and  Hamiltonian (yellow) limits. 
As expected \cite{gautier2022}, the dissipative confinement shows an exponential suppression of $\Gamma$ significantly outperforming its Kerr counterpart.
We have verified that, if we set $\Delta=0$ and optimize $\Gamma$ over $\theta$ alone, the optimal point coincides with the dissipative limit ($\theta_{\rm opt}|_{\Delta=0}=0$). We thus move to consider the effect of $\Delta$.
By setting $\theta=0$ and optimizing over $\Delta$ alone, we find that $\Delta_{\rm opt}|_{\theta=0}=0$. In the dissipative limit, therefore, there is no advantage in introducing a nonvanishing detuning.
In the Kerr limit ($U/\eta=10^{6}$) instead, optimizing $\Delta$ introduces only a marginal advantage in the dephasing-dominated configuration [Fig.~\ref{fig:gamma_v_G}(a)], but a significant one in that dominated by photon-loss [Fig.~\ref{fig:gamma_v_G}(b)].
Finally, we consider the combined optimization of $\theta$ and $\Delta$.
We verify that the critical cat code significantly outperforms its resonant counterparts, and confirm that in the dephasing-dominated case, for all $G>3$, the optimum is found deep in the hybrid regime ($U/\eta \approx 10^2$), while in the photon-loss dominated one, it approaches the Kerr limit ($U/\eta \gtrsim 10^5$).

\begin{figure}[tb]
\center
\includegraphics[width=0.49\textwidth]{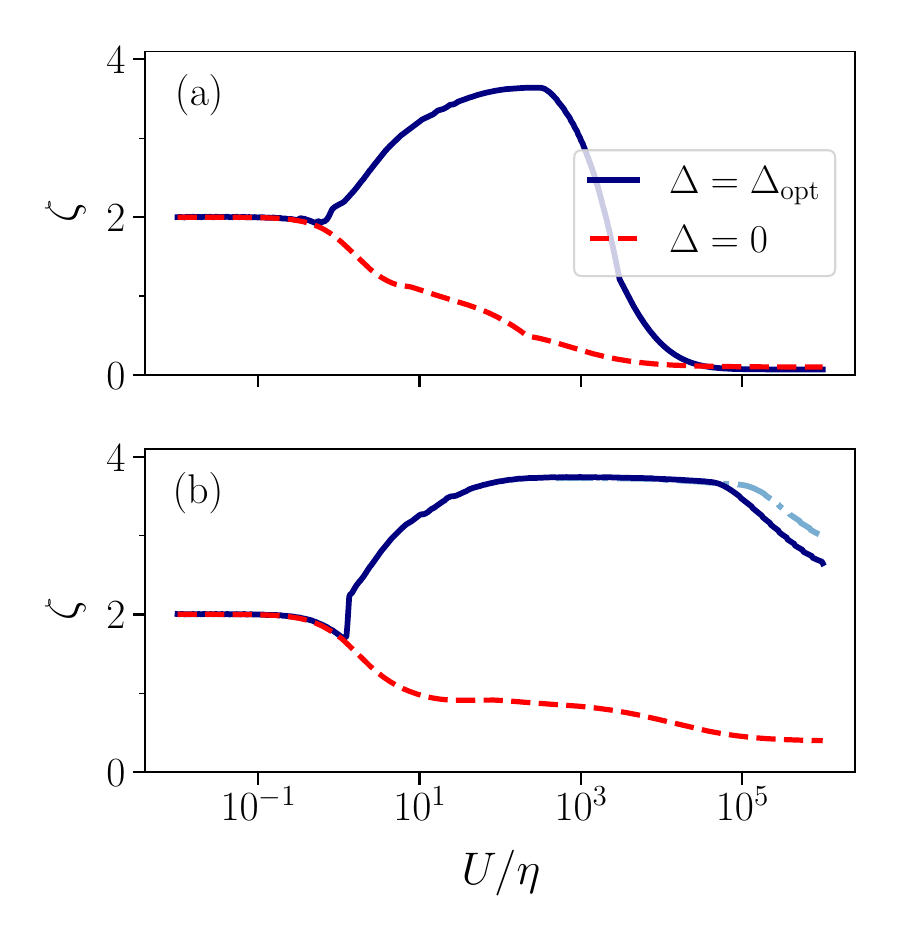}
\caption{\label{fig:scaling_coeff} The exponential rate $\zeta$ computed as a function of $U/\eta$ for $\Delta=0$ (dashed) and $\Delta=\Delta_\mathrm{opt}$ (solid). (a) $\kappa_1 = 10^{-5}\ll \kappa_\phi= 10^{-3}$; (b) $\kappa_1 = 10^{-3}\gg \kappa_\phi= 10^{-5}$. The dot-dashed curve in panel (b) shows $\zeta(\Delta_{\rm opt})$ for $\Delta_{\rm max} = 90$.}
\end{figure}

We now set out to demonstrate the enhanced scaling of $\Gamma$ as a function of the photon number in the critical cat.
We find that in the driving range $4<G<12$, all curves display the functional dependence $\Gamma/G \simeq \Gamma_0 \exp{- \zeta G}$, which we use to extract the scaling coefficient $\zeta(\theta,\kappa_1,\kappa_\phi)$. 
The results are shown in Fig.~\ref{fig:scaling_coeff} as a function of the ratio $U/\eta$. As also predicted in Ref.~\cite{gautier2022}, for $\Delta=0$ the system progressively loses its protection against dephasing-induced bit-flip errors as it transitions from the dissipative to the Hamiltonian limit. On the contrary, operating the critical cat at $\Delta = \Delta_{\rm opt}(G)$  endows the system with an additional resilience to errors, so that the bit-flip error rate is not only exponentially suppressed in the photon number, but it does so with an enhanced scaling coefficient compared to its optimal resonant value $\zeta=2$ achieved in the purely dissipative regime.
To clarify the relevance of the bound $\Delta_{\rm max}$, Fig.~\ref{fig:scaling_coeff}(b) also shows the rate $\zeta$ obtained by setting $\Delta_{\rm max}=90$. This larger bound results in a marginal increase of the favorable range of values of $U/\eta$, as also expected from the analysis of $\Delta_{\rm opt}(\theta)$ in Fig.~\ref{fig:gamma_v_Delta_v_U}.
We finally draw attention to the fact that in a realistic framework, where detuning is a limited resource, the presence of a small two-photon loss is essential to achieve optimal scaling, and operating with a pure Kerr confinement can result in a severe under-performance of the cat.

Overall, we demonstrate the existence of regions in parameter space granting enhanced protection from bit-flip errors. Our findings also resolve the issues tied to the hybrid operation of Kerr cat qubits outlined in Ref.~\cite{gautier2022} and discussed in Section \ref{sec:intro}. Finally, we remark that in this section we have treated $U/\eta$ as as free parameter, though in general $U$ and $\eta$  represent distinct and limited resources. 
Nonetheless, we have shown that the optimal configuration is one where $\eta \sim \kappa_1, \, \kappa_{\phi} \ll U$. 
It is thus possible to achieve these values starting from a Kerr qubit architecture through reservoir engineering techniques, as demonstrated in previous works \cite{grimm2020, leghtas2015}.

\section{Metastable encoding and the critical cat}
\label{sec:encoding}

The analysis until now focused on the minimization of $\Gamma$ alone. We now consider the interplay between $\Gamma$ and $\Gamma_{\rm leak}$ and the various possible encodings.

\begin{figure}[htb]
\includegraphics[width=0.48\textwidth]{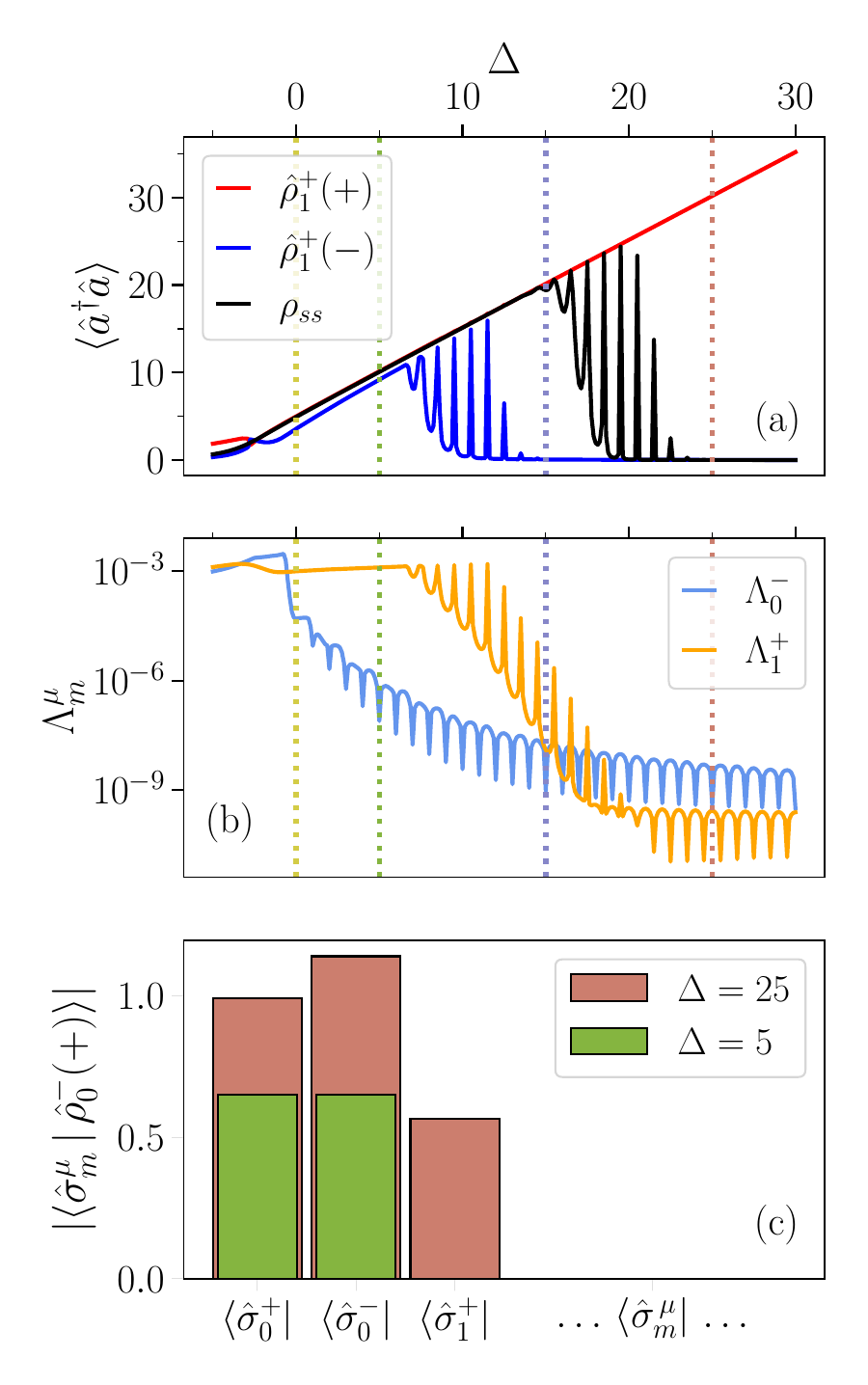} 
\caption{\label{fig:deltascan} (a) Average photon number of the steady and metastable states as defined in the main text. The jump discontinuity in the photon number signals the presence of a first-order DPT at $\Delta_c\sim15$ \cite{Note2}. 
(b) Liouvillian gaps $\Lambda_1^+$ and $\Lambda_0^-$ as defined in Eq.~\eqref{Eq:block_eigenvalues_weak}. 
Vertical dotted lines denote the values of $\Delta$ selected for further investigation in Fig.~\ref{fig:preparation}. (c) Expansion of the state $\hat\rho_0^-(+) \equiv\ketbra{1_L}$ [cfr Eq.~\eqref{eqn:spec_decomp}] in the left eigenstates of $\LL$. 
The expansion coefficients  are $c_m = \braket{\sigma^{\mu}_m}{\hat\rho_0^-(+)}$.
Each component in the expansion evolves with one specific timescale.
For $\Delta<\Delta_c$, only two components are nonvanishing: the steady state and that corresponding to bit flips. 
For $\Delta>\Delta_c$, a single additional component corresponding to the leakage with rate $\Gamma_{\rm leak} = \Lambda_1^+$ emerges, and the dynamics is that of Eq.~\eqref{Eq:metastable_encoding}. We consider $G=5$, $U/\eta = 10^{5}$, and $\kappa_1=10^{-3}\gg10^{-5}=\kappa_\phi$.}
\end{figure}

A fundamental property of the two-photon driven-dissipative Kerr resonator is the occurrence of a first-order DPT at a critical detuning $\Delta_c$, resulting in a discontinuous change in the steady state of the system \cite{bartolo2016, casteels2017, fink2018}. 
Here, we show that this phase transition has profound consequences on the encoding and determines the way the qubit must be prepared and controlled to achieve optimal performance.
The DPT arising for a specific choice of parameters is shown in Fig.~\ref{fig:deltascan}(a). Here, the system's photon number is seen to change abruptly as $\Delta$ is increased beyond $\Delta_c$ \footnote{At finite driving strength, only the finite-size precursors of an actual phase transition are witnessed \cite{minganti2018, rota2017, biella2017}.
Here, the critical region encompasses a finite range of detunings where the system crosses over from the highly-populated phase to the other, by showing a series of peaks in the photon number. 
We define $\Delta_c$ as the median value of this critical region.
}. 
Indeed, while for $\Delta<\Delta_c$ the system's steady state defines a highly-populated phase (cat-like), for $\Delta>\Delta_c$ the steady state approximates the squeezed vacuum state.

Figures \ref{fig:gamma_v_Delta_v_U}(a)~and~(b) detail the dependence of $\Delta_{c}$ on $\theta$, marking the boundary between the two phases. 
This figure shows that $\Delta_\mathrm{opt}$ minimizing $\Gamma$ crosses $\Delta_c$ as the ratio $U/\eta$ is increased.
A fundamental question is therefore which kind of encoding among those introduced above holds in each phase.
Another main result of our work consists in showing that the two phases identified above express different encodings, with the metastable encoding in Eq.~\eqref{Eq:metastable_encoding} characterizing all largely-detuned configurations.
This is a pivotal point, as now $\Gamma_{\rm leak}$ also contributes to the performance of the code. This effect has been overlooked when modeling the cat code within a purely Hamiltonian formalism, oblivious to the DPT.
We will also show that the critical behaviour affects the initialization of the code in either phase, and we will provide an initialization protocol that prevents the system from being pinned to the squeezed vacuum.
All in all, accounting for the critical behaviour of the cat-code at finite values of $\Delta$, $\eta$, $\kappa_1$, and $\kappa_{\phi}$ is  key to determine an optimal encoding protocol.

The analysis that follows assumes $G=5$, $U/\eta = 10^{5}$, and $\kappa_1=10^{-3}\gg10^{-5}=\kappa_\phi$.

\begin{figure}[htb]
\includegraphics[width=0.48\textwidth]{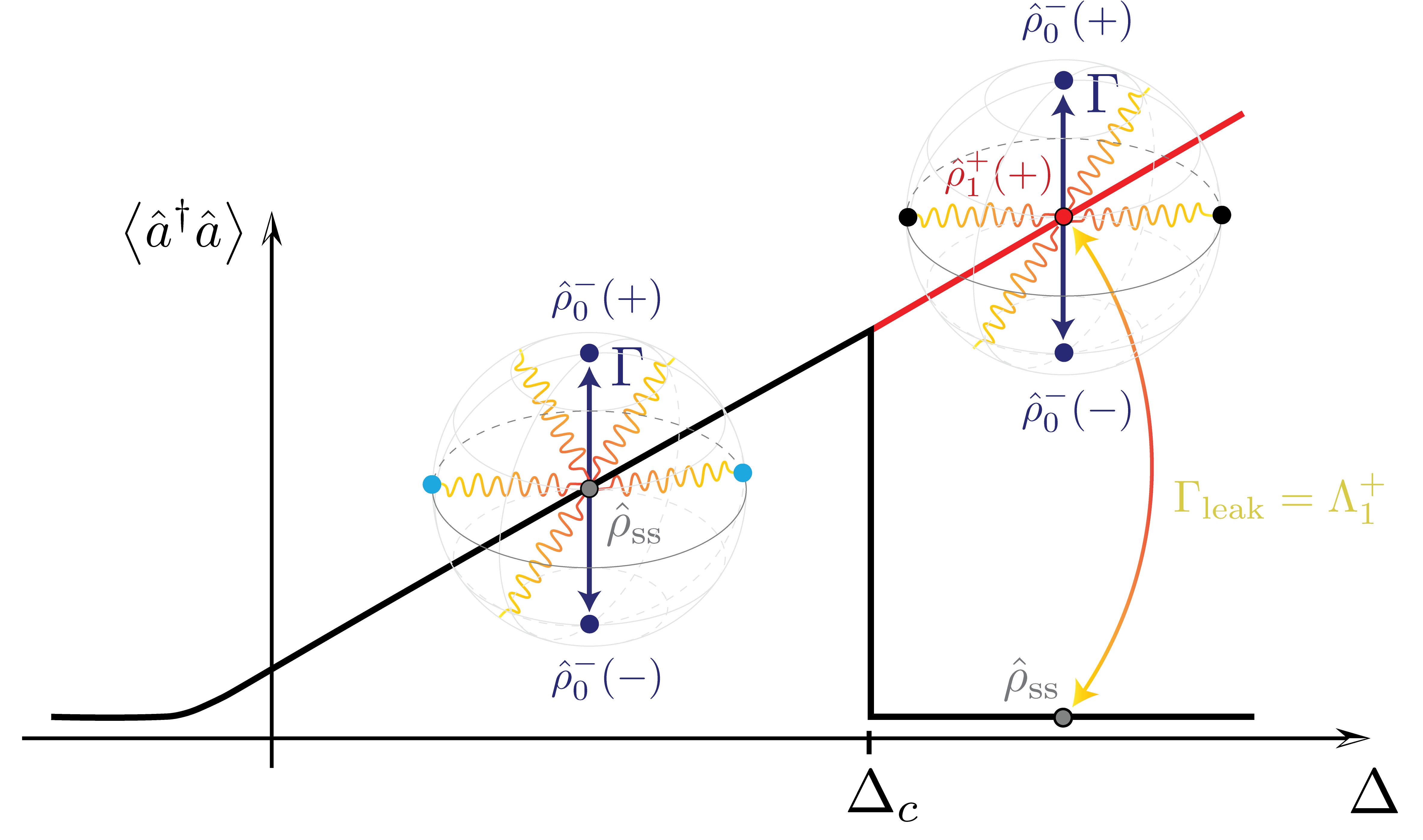} 
\caption{\label{fig:metascheme} Illustration of the optimal encoding on the two sides of the phase boundary. For $\Delta<\Delta_c$ the steady-state manifold coincides with the encoding manifold, and the logical code space is characterized by a single phase-flip error rate $\Gamma$. For $\Delta>\Delta_c$ the steady state is a squeezed vacuum and is distinct from the optimal encoding manifold characterized by a finite average photon number. In this case, a second timescale emerges, which is characterized by the rate $\Lambda_1^+$ at which the logical states decay into the vacuum.}
\end{figure}

\subsection{Criticality and metastability}

Two types of critical phenomena can occur in open quantum systems: first- and second-order phase transitions.
The former (latter) corresponds to a discontinuous (continuous, but not differentiable) change in the system's properties, whose finite-size precursor is shown in Fig.~\ref{fig:deltascan}(a), displaying the steady-state photon number as a function of $\Delta$. 
Upon crossing the critical value $\Delta_c$, an abrupt transition from a highly-populated to a squeezed vacuum phase (hereunder, vacuum for brevity) occurs. 
We verified that $\Delta_c$ is only marginally affected by the values of $(\kappa_1, \kappa_\phi)$.
The hallmark of a first-order DPT is the closure of the Liouvillian gap $\Lambda_1^{+}$, whose finite-size precursor is shown in Fig.~\ref{fig:deltascan}(b). 
This closure entails metastability and is associated with a hysteretical behaviour of the system which, despite the uniqueness of the steady state, can persist in other states for very long times.

In the cat code, a biased response to noise requires the existence of two opposite points on the logical Bloch sphere between which noise-induced transitions are exponentially suppressed. 
Without loss of generality, we can identify these two points as the logical states $\ket{0_L}$ and $\ket{1_L}$. 
These states are defined through the following procedure \cite{rivas2012, minganti2018}. 
We start from the eigenoperator $\hat\rho_0^-$, that in its diagonal form reads
\begin{equation}\label{Eq:diagonalizaion_DM}
    \hat\rho_0^- = \sum_{i}p_i\ketbra{\psi_i}{\psi_i}.
\end{equation}
Because $\hat\rho_0^-$ is Hermitian, all the $p_i$ are real, and $\braket{\psi_i}{\psi_j}=\delta_{ij}$. Since $\Tr[\hat \rho_0^-]=0$, some $p_i$ will be positive while others negative and we can order them in such a way to have $p_i > 0$ ($p_i < 0$) for $i \leq  \bar i$ ($i > \bar i$). In this way $\hat\rho_0^- = \hat\rho_0^{-}(+) - \hat\rho_0^{-}(-)$ with 
\begin{equation}
\label{eqn:spec_decomp}
    \hat\rho_0^{-}(+) = \sum_{i\leq \bar i} p_i\ketbra{\psi_i},\,\,\, \hat\rho_0^{-}(-) = -\sum_{i>\bar i}p_i\ketbra{\psi_i}.
\end{equation}
The probabilities $p_i$ are normalized to ensure that $\Tr [\hat \rho_0^{-}(\pm)]= 1$.
The logical states of the code are then defined as
\begin{equation}
    \ketbra{0_L} = \hat\rho_0^{-}(+), \quad \ketbra{1_L} = \hat\rho_0^{-}(-)
\end{equation}
This definition ensures that the bit-flip rate 
$\Gamma = \Lambda_0^{-}$ coincides with the slowest decay rate of the off-diagonal Liouvillian sector.
Additionally, for all cases considered below, the logical qubit will undergo decoherence at a rate $\Gamma_{\phi}$ -- the noise mechanism not suppressed by the cat encoding.
\\

We identify three regimes:
\begin{description}
\item[$\mathbf{\Delta=0}$ ] Here $\hat\rho_0^- \propto \ketbra{\alpha} - \ketbra{-\alpha}$, thus $\ket{0_L} = \ket{\alpha}$ and 
$\ket{1_L} = \ket{-\alpha}$.
The very same states also define $\hat\rho_{\rm ss} \simeq (\ketbra{\alpha} + \ketbra{-\alpha})/2$, which is located at the center at the Bloch sphere (see illustration in Fig.~\ref{fig:metascheme}).
This is the steady-state encoding defined in Eq.~\eqref{Eq:statedy-state_encoding} for which the states $\ket{0_L}$ and $\ket{1_L}$ are \textit{only} exchanged at a rate $\Gamma$.

\item[$\mathbf{0<\Delta<\Delta_c}$ ] 
While $\hat\rho_0^-$ cannot be expressed as a simple mixture of two coherent states, as in the case when $\Delta=0$, a steady-state encoding still holds (as illustrated in Fig.~\ref{fig:metascheme}) since $\sss \simeq \ketbra{0_L} + \ketbra{1_L}$. The spectral decomposition in Fig.\ref{fig:deltascan}(c) demonstrates that also in this regime $\ket{0_L}$ and $\ket{1_L}$ \textit{only} flip at a unique rate $\Gamma$.
As $\Delta$ approaches $\Delta_c$, the vacuum state becomes progressively more stable, as shown by the decrease of $\Lambda_1^+$ in Fig.~\ref{fig:deltascan}(b). This increased stability makes the initialization of the code more challenging as the system tends to remain pinned to the long-lived vacuum state.

\item[$\mathbf{\Delta>\Delta_c}$ ] In this regime,
$\sss \simeq \ketbra{n=0}{n=0} \neq \ketbra{0_L} + \ketbra{1_L}$.
This is a metastable encoding according to Eq.~\eqref{Eq:metastable_encoding}.
Indeed, all processes within the code space can still be described in terms of the logical error rates $\Gamma$ and $\Gamma_{\phi}$.  All states initialized on the logical Bloch sphere will however irreversibly decay towards $\sss$, which now lies outside of the code space. 
This process occurs at a rate $\Lambda_1^+ = \Gamma_{\rm leak}$.
The center of the Bloch sphere, which before coincided with $\sss$, is now $\hat\rho_1^{+}(+)$ and is obtained by the spectral decomposition
\begin{equation}
\label{Eq:eigen_rho_1_+}
    \hat\rho_1^{+} = \hat\rho_1^{+}(+) - \hat\rho_1^{+}(-),
\end{equation}
defined in analogy with Eqs.~\eqref{Eq:diagonalizaion_DM}~and~\eqref{eqn:spec_decomp}.
Indeed, while $\hat\rho_1^{+}(-)$ closely resembles the vacuum-like steady state, $\hat\rho_1^{+}(+) \propto \ketbra{0_L} + \ketbra{1_L}$ (see Fig.\ref{fig:MS_schematic}).
In this regime the quality of the encoding will thus depend not only on the magnitude of $\Gamma$, $\Gamma_{\phi}$, but on that of $\Gamma_{\rm leak}$ as well.
Our simulation suggest that, for $\Delta_{\rm opt}>\Delta_c$, the rates $\Gamma$ and $\Gamma_{\rm leak}$ may have comparable magnitudes, as can be inferred from the data in Fig.~\ref{fig:deltascan}(b).

\end{description}

\begin{figure}[htb]
\includegraphics[width=0.49\textwidth]{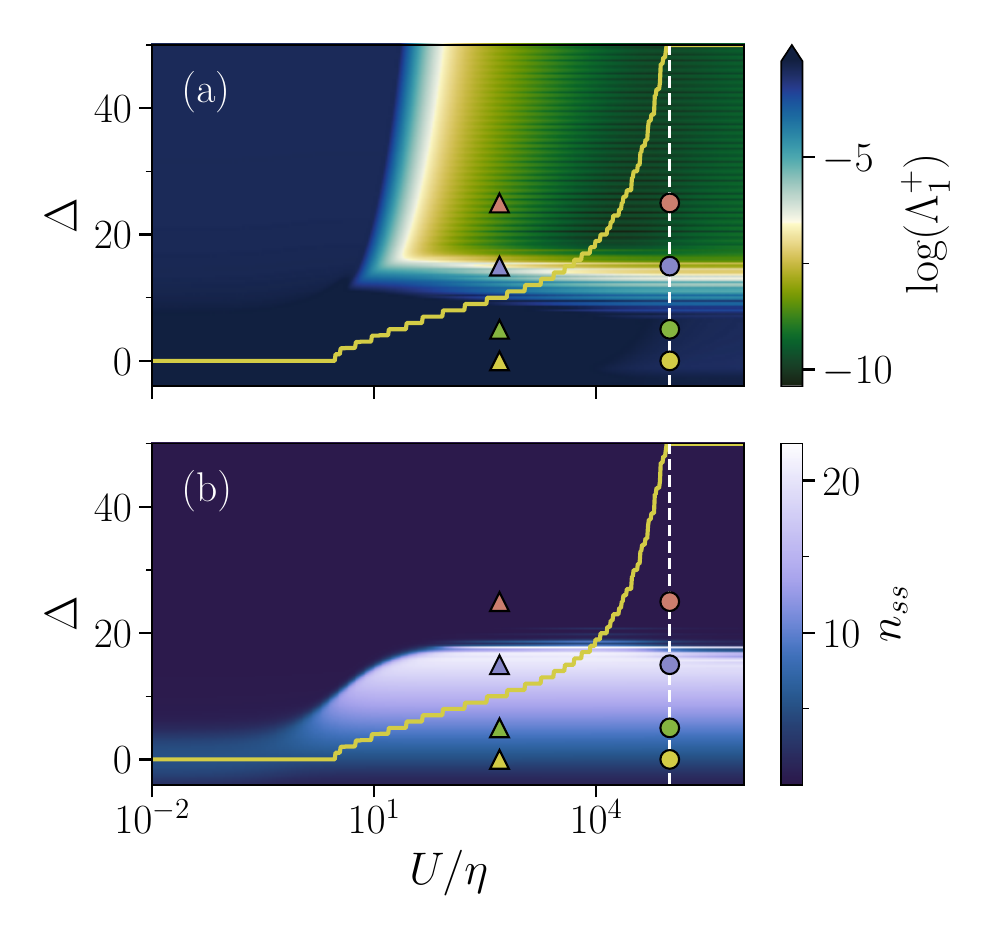} 
\caption{\label{fig:diagonalgap} Phase diagram of the diagonal Liouvillian gap $\Lambda_1^+$ and steady-state photon number (b) as a function of detuning $\Delta$ and the ratio $U/\eta$. The yellow line denotes the optimal detuning $\Delta_\mathrm{opt}$ that minimizes the bit-flip error rate $\Gamma$.
The circles and triangles indicate specific the cases selected for the discussion of the state preparation protocol in
Sec.~\ref{Sec:preparation_protocol} and shown in Fig.~\ref{fig:preparation}.}
\end{figure}

We remark that both $\Lambda_0^{-}$ and $\Lambda_1^{+}$ display oscillations with sharp minima.
While the minima of $\Lambda_0^{-}$ occur at $\Delta \simeq U m$ with $m\in\mathbb{N}$, those of $\Lambda_1^{+}$ occur at $\Delta = (m+1/2) U$, where $\Lambda_0^{-}$ has maxima.
While in the regime where a steady-state encoding holds $\Lambda_1^{+}$ does not affect the code's performance, in the one characterized by a metastable encoding, $\Lambda_1^{+} = \Gamma_{\rm leak}$ defines a source of uncorrectable errors. Because for the detuning values minimizing $\Lambda_0^{-}$, $\Lambda_0^{-}$ and $\Lambda_1^{+}$ become comparable, a trade-off between these two error processes emerges when deciding at which detuning to operate the code.
The conclusions drawn here hold in both noise configurations, and throughout the range $10^{2}\leq U/\eta \leq 10^{5}$. 
Additional results in this direction are presented in Appendix~\ref{sec:app_extension}. 
From these data,  we notice that these oscillatory features become less pronounced the larger $\eta$ we take.

In Fig.~\ref{fig:diagonalgap}, we finally display the Liouvillian gap $\Lambda_1^+$ and the steady-state average photon number as a function of $\Delta$ and $U/\eta$. 
As $U/\eta$ increases, the optimal detuning $\Delta_\mathrm{opt}$ crosses the phase boundary, identifying large regions in the parameter space where the optimal encoding is the metastable one.
In all parameters considered, the sudden drop in the photon number is accompanied by a reduction in $\Lambda_1^+$, minimal around $\Delta_{\rm opt}$. 
Comparing Figs.~\ref{fig:diagonalgap}(a)~and~\ref{fig:gamma_v_Delta_v_U}(b) we conclude that a minimal leakage error $\Gamma_{\rm leak}$ accompanies the minimal  $\Gamma$.
These data support the analysis above, demonstrating that the presence of a metastable encoding is an emergent feature of the cat code at finite detuning and highlighting the key role of criticality in determining the optimal encoding.

\begin{figure}[htb]
\includegraphics[width=0.48\textwidth]{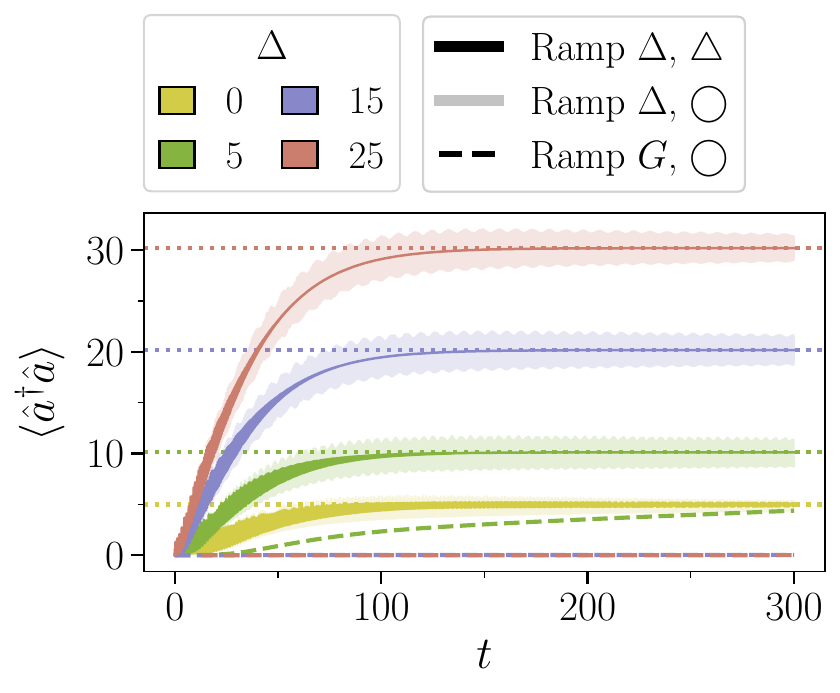} 
\caption{\label{fig:preparation}  Time dependent simulation of the initialization protocols detailed in the main text. Dotted horizontal lines define the desired photon number for the different detuning configurations. Colors and symbols in the legend correspond to those points selected in the phase diagram of Fig.~\ref{fig:diagonalgap}. Dashed lines represent the adiabatic ramping of $G(t) = f(t,\tau) G$ where $f(t,\tau) = \tanh(t/\tau)$, with $\tau=50$ and $G=5$.
Solid lines describe $\Delta(t) = f(t,\tau) \Delta$ in the presence of two-photon loss. 
The shaded lines describe the amplitude of the oscillation of the same $\Delta(t)$ protocol, setting $U/\eta = 10^{5}$.
}
\end{figure}

\subsection{State preparation protocol}
\label{Sec:preparation_protocol}

The scenario described above indicates that the code space is stable and the vacuum metastable whenever $\Delta<\Delta_c$, while the converse is true for $\Delta>\Delta_c$.
Assuming a preparation protocol where, starting from the vacuum, the two-photon driving field $G$ is ramped up, two issues may arise.
For $\Delta<\Delta_c$, the system will remain stuck in the vacuum for a time $1/\Lambda_0^{-}$ before reaching the stable code space.
For $\Delta>\Delta_c$, the system will never leave the stable vacuum and thus never reach the metastable code space.
We simulate this simple ramp-up protocol in Fig.~\ref{fig:preparation}, confirming the predictions above.

We propose a solution for this initialization problem taking advantage of criticality. 
While the system undergoes a first-order DPT for $\Delta>0$, a second-order DPT takes place at $\Delta<0$ \cite{savona2017,bartolo2016,minganti2018}. 
As shown in Ref.~\cite{dicandia2021, ilias2022, fernandez-lorenzo2017}, around a second-order DPT, the system is highly responsive and does not experience critical slowing down if properly initialized. 
We take advantage of this property in a protocol where $\Delta$ is ramped up rather than $G$.
We start from $\Delta(t=0)<G$, the system in the vacuum state, and $G$ fixed to its target value.
We then increase $\Delta(t)$ according to a schedule
\begin{equation}
    \Delta(t) = f(t, \tau) \, \Delta,
\end{equation}
where $\Delta$ is the target detuning value, and $f(t, \tau)$ a smooth function increasing from $ f(0, \tau) = 0$ to $f(t\to \infty, \tau) =1$ over a characteristic time $\tau$.
For very small values of $\eta$ this protocol leads to the desired state manifold, but it is characterized by unwanted fast oscillations around the target state.
These are due to the phase accumulation through the adiabatic sweep, and hinder the exact initialization of the code.
In Fig.~\ref{fig:preparation} we show a slightly larger value of $\eta$ suppresses these unwanted oscillations.
This choice does not significantly increase the rate $\Gamma$ from its optimal value [c.f.~, e.g., Fig.~\ref{fig:gamma_v_Delta_v_U}(b)].
This result further advocates in favour of a hybrid operation of the cat instead of the Kerr limit, even in photon-loss dominated configurations.

\section{Resistance to frequency shifts}
\label{sec:Frequency_shift}

As discussed in Ref.~\cite{lieu2020, chamberland2022}, random shifts in the resonant frequency of the oscillator can significantly hinder the code's performance. 
In most cases, these effects can be modeled by an additional correlated noise in the form of an effective detuning $\Delta_{\mathrm{err}} \a^\dagger \a$ acting on the system for a limited time $t_0$, representing, e.g., the typical time of a gate operation(s).
Common sources of shifts are cross-Kerr interactions originating either from spurious interactions with dissipatively coupled qubits, or from stochastic jumps and thermal excitations in reservoir modes nonlinearly coupled to the system by, e.g., Josephson junctions.
Because quantum information can be efficiently encoded over the whole broken symmetry region as in that realizing the metastable encoding, we anticipate the effect of such shifts to be far less detrimental for the critical cat code.

To demonstrate this claim, we consider the following protocol. The system is initialized in the state $\hat{\rho}(0)=\ketbra{0_L}{0_L}$ and subsequently quenched with an additional detuning $\Delta_{\mathrm{err}}$. The system evolves under $\LL^\prime \hat{\rho} = \LL_0  \hat{\rho} -i \Delta_{\mathrm{err}} [\a^\dagger \a,  \hat{\rho}]$ for a time $t_0=10$ at which point $\Delta_{\rm err}$ is switched off, leaving the the system to stationarize under $\LL_0$ over a time $t_1 = 5$.

\begin{figure}[htb]
\includegraphics[width=0.48\textwidth]{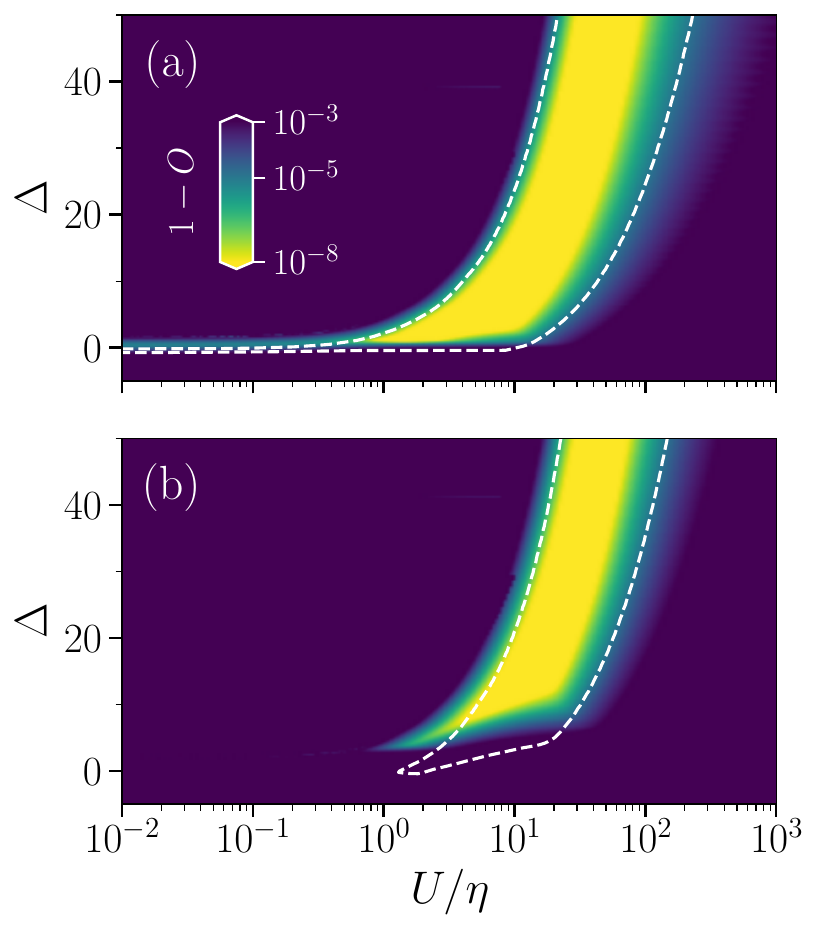} 
\caption{\label{fig:shifts} 
The quantity $1-O$, defined in Eq.~\eqref{eqn:overlap}, displayed as a function of the detuning $\Delta$ and of the angle $\theta$ in Eq.~\eqref{Eq:W_definition} for $\kappa_1 = 10^{-3}\gg \kappa_\phi= 10^{-5}$. The data is obtained by appropriately selecting at each point the correct encoding, either steady-state or metastable.
The initial state $\ketbra{0_L}$, and the final state obtained from the recovery procedure, are detailed in the main text. 
The data in the two panels were computed assuming a frequency shift of (a) $\Delta_{\rm err} = 1$ and (b) $\Delta_{\rm err} = 4 $.
The white dashed lines bound the region wherein $1 − O < 10^{−5}$  in the cases (a) $\Delta_{\rm err} = - 1$ and (b) $\Delta_{\rm err} = -4 $. 
}
\end{figure}

Figure~\ref{fig:shifts} shows the orthogonality $1-O$ as a function of $\Delta$ and $U/\eta$. 
$O$ is the overlap between the initial and the final states, it is defined by \footnote{We choose the overlap to quantify the resistance of critical cat because it is more numerically stable than other indicator (e.g., the fidelity). Qualitatively analogous results are found using the fidelity measure for the distance between the initial and final states.}
\begin{equation}
\begin{aligned}
\label{eqn:overlap}
    O&\equiv\Tr[\hat{\rho}(0)\, \hat{\rho}(t_0 +t_1)] 
    = \Tr[\hat{\rho}(0)\, e^{\LL_0 t_1} e^{\LL' t_0} \hat{\rho}(0) ],
\end{aligned}
\end{equation} 
and can be related to the evolution of the observables characterizing the system using the asymptotic projection method \cite{lieu2020}.
Figure~\ref{fig:shifts}(a) is computed for $\Delta_{\rm err} = 1$. We choose this value in accordance with the discussion in App.~B3 of Ref.~\cite{chamberland2022}, specializing its result to the case of two coupled oscillators. The physical parameters characterizing storage and reservoir modes were chosen by interpolating between those found in Refs.~\cite{grimm2020, leghtas2015, touzard2018, frattini2022}.
The data clearly identify a range of parameters in which the quantum information encoded in the critical cat is highly resilient to random frequency shifts. This region is characterized by positive values of $\Delta$ and $ 1 \lesssim U/\eta\lesssim 10^3$.
A similar analysis for $\Delta_{\rm err} = 4 $ is shown in Fig.~\ref{fig:shifts}(b). 
This shows that even for very large frequency shifts, the critical cat offers still room for optimal encoding.
This analysis further demonstrates the advantage of a hybrid encoding.
We however notice that the parameters granting optimal protection from frequency shifts do not coincide with those minimizing $\Gamma$.
Nonetheless, a tradeoff between these two optimal conditions can be found, but we anticipate the necessity for a complete characterization of these effects in coupled cat qubits when designing two-qubit gates or concatenation protocols, as also discussed in Ref.~\cite{chamberland2022}.

\begin{figure}[b]
\includegraphics[width=0.48\textwidth]{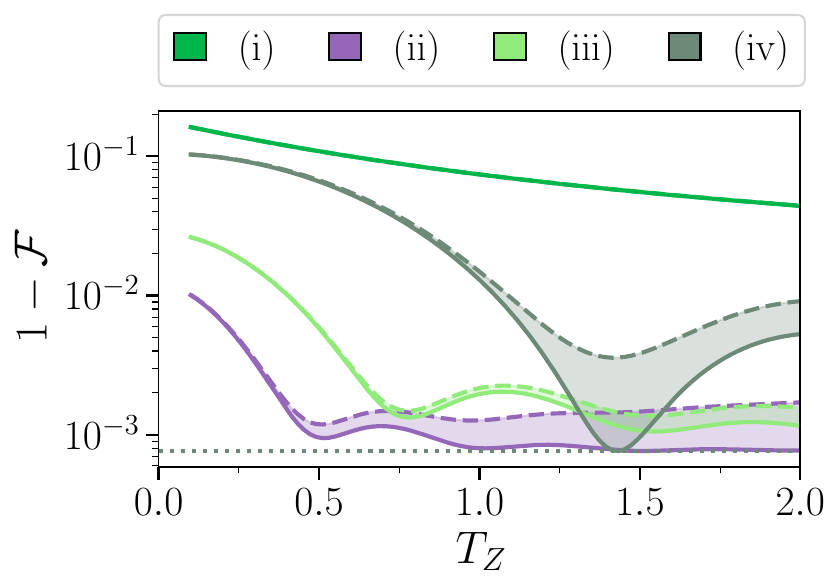} 
\caption{\label{fig:Zgate} Gate infidelity as defined by Eq.~\eqref{eqn:gate_inf}. We consider four different operation regimes (i):~$\Delta=0$, $U/\eta=0$ (resonant-dissipative); (ii):~$\Delta=30$, $U/\eta=50$; (iii):~$\Delta=10$, $U/\eta=50$ and (iv):~$\Delta=0$, $U/\eta=10^6$ (resonant-Hamiltonian). Solid and dashed lines correspond to applying the gate at $t=0$ or after an idling time of $t=2T_Z$.
We consider the noise configuration $\kappa_1 = 10^{-5}\ll \kappa_\phi= 10^{-3}$.}
\end{figure}

\section{Bias-preserving gates}
\label{sec:gates}
To harness the full potential of biased-noise qubits, and retain their benefits throughout the computation, gate operations must preserve the exponential suppression of bit-flip errors. Possible implementations of these \emph{bias-preserving gates} rely on Zeno dynamics and topological deformation of the code space, and have been extensively investigated in the resonant-dissipative and resonant-Hamiltonian configurations \cite{puri2020, guillaud2021, chamberland2022, guillaud2019, mirrahimi2014, gautier2022}. The very same protocols can be applied in the presence of non-vanishing detuning and in the hybrid-regime. This possibility has been extensively investigated in Refs.~\cite{gautier2022, ruiz2022}, also in relation to colored dissipation and super-adiabatic pulse designs. To complete our investigation we provide evidence of analogous benefits stemming from operating the qubit in the hybrid-detuned regime. We do so by investigating the simple example of the bias-preserving realization of single-qubit $Z$-gates via $\H_{Z} = F(\a + \a^\dagger)$, with $F$ a weak drive.
Indeed, $\ketbra{+_L}{+_L}$ and $\ketbra{-_L}{-_L}$ are states of opposite parity, which can be connected by single-photon exchange. The analytical expression for the effective frequency (period) $\Omega_Z = 2F\Re{\alpha}$ ($T_Z = \pi/\Omega_Z$) of the Rabi oscillation between these two states is determined by neglecting the effect of errors, setting $\Delta=0$, and evaluating the action of $\H_Z$ onto the manifold spanned by the pure cat states $\ket{{\cal C}^{\pm}_\alpha}$ with $\alpha=\sqrt{G/W}$.
In the detuned regime, we find a similar expression for $\Omega_Z$ with $\alpha = \sqrt{\langle{\hat{a}^2} \rangle}$. 
The optimal driving strength $F$ maximizing the performance of the gate is determined by the competition between the adiabatic and non-adiabatic errors induced by the gate operation. Adiabatic errors are those caused by the single-photon loss and dephasing events occurring during the action of the gate. Non-adiabatic errors, on the other hand, stem from diffusion out of the code manifold induced by the driving field. Under these conditions, a trade-off must be fund between these two sources of error.

Figure~\ref{fig:Zgate} provides a quantitative comparison of the performance of $Z$-gates in the resonant-dissipative (i), hybrid-detuned (ii,iii), and resonant-Hamiltonian (iv) configurations. The simulation is performed by initializing the system in $\ketbra{+_L}$ and evolving under $\LL_Z\hat\rho = \LL\hat\rho -i[\H_Z,\hat\rho]$ for a time $T_Z$. The gate infidelity is defined as  
\begin{equation}
    \label{eqn:gate_inf}
    1-\mathcal{F} = \frac{\expval*{\hat \Pi}|_{T_Z/2}+1}{2},
\end{equation} 
exploiting the fact that after half a Rabi oscillation the system's state should coincide with $\ketbra{-_L}$, and that $\operatorname{Tr}[\hat \Pi \ketbra{\pm_L}]=\pm1$, where $\hat\Pi$ is the parity.  
Our results confirm the viability of performing high-fidelity gates in the hybrid-detuned configuration, and show a substantial improvement in the speed of gate operation in the critical regime with minimal impact on its fidelity. 

In realistic computation, however, one needs also to asses how the gate performs when acting on an initial state deformed by the spurious processes of prior operations.
In this paper we consider the minimal example of the $Z$-gate being applied after an idling time twice as large as $T_Z$.
The results, displayed in Fig.~\ref{fig:Zgate}, clearly demonstrate the advantage of the critical encoding over both the dissipative and Hamiltonian ones.

\section{Discussion and conclusions}
\label{Sec:Conclusions}

We have investigated the properties of the Schrödinger's cat code in the whole range of regimes between the two limiting cases of dissipative \cite{leghtas2015,mirrahimi2014,touzard2018} and Kerr \cite{puri2017,puri2019,puri2020,grimm2020} cat code. We have demonstrated that operating the cat code at finite values of the detuning between the two-photon driving field and the cavity frequency dramatically increases the resilience of the cat qubit to bit-flip errors. 
This improvement may require to encode the system in the metastable manifold emerging from the first-order dissipative phase transition characterizing the driven-dissipative Kerr resonator at finite detuning \cite{bartolo2016}. 
We have also discussed how one needs to take into account several factors when discussing the performance of detuned cat codes.
Beyond the phase flip errors $\Gamma$, one has to explicitly take into account the fact that the system is metastable in many configurations, and therefore a new error process, characterized by a rate $\Gamma_{\rm leak}$, needs to be considered when discussing the quality of the encoding.

Our analysis of the bit-flip rate demonstrates that a small amount of two-photon loss already produces a significant improvement of the code performance with respect to the Kerr limit.
Furthermore, we propose an efficient initialization protocol relying on small two-photon loss that circumvents the problems posed by the hysteresis of the vacuum and the metastable nature of the encoding.
Two-photon loss is also beneficial for correcting frequency-shift errors that are expected to emerge in connected and concatenated cat architectures, as detailed in Ref.~\cite{chamberland2022}. 
Finally, we show that bias-preserving gates perform significantly better in the hybrid-detuned case that in Kerr and dissipative limits.

The picture emerging from our analysis is that focusing on a single a figure of merit to choose the optimal regime operation is reductive.
All quantities considered here, namely $\Gamma$, $\Gamma_{\rm leak}$, the resistance to frequency shifts, code initialization, and gate fidelity, advocate for a small, but non-negligible, two-photon loss rate being necessary for the optimal performance of the code.
However, these quantities do not identify a unique global optimum. The optimal choice for the the two-photon loss rate will ultimately depend on the specific of the platform under consideration. 

All in all, dissipation and detuning emerge as pivotal and necessary resources for optimally efficient and reliable bosonic quantum encoding.
From a broader perspective, our work demonstrates the nontrivial nature of dissipative processes in the presence of criticality.
It suggests that the largely unexplored parameter space of all bosonic codes may still offer regions where specific properties of the code are enhanced, leading to a competitively efficient design of bosonic quantum code architectures.

\begin{acknowledgments}
We acknowledge useful discussions with Victor V. Albert, Alexander Grimm, Carlos S\'anchez-Mu\~noz, and David S. Schlegel. 
This work was supported by the Swiss National Science Foundation through Project No. 200020\_185015, and was conducted with the financial support of the EPFL Science Seed Fund 2021.
\end{acknowledgments}

\appendix

\section{Bosonic quantum information encoding}
\label{app:encoding}
In this section we focus on demonstrating the validity of the encoding for the different configurations explored in the main text. 
To do so, we first need to assess under which conditions a set of density matrices can encode quantum information and allows for quantum computation.
Consider the six generic matrices $\hat{\rho}_{\pm X}, \, \hat{\rho}_{\pm Y}, \, \hat{\rho}_{\pm Z}$.
The first requirement for them to define a logical Bloch sphere is that the states on the opposite sides of the Bloch sphere are pairwise orthogonal. 
Having verified this, for them to support gate operations, one must also ensure the structure they generate to be isomorphic to
\begin{equation}
\label{Eq:NS}
\hat{\rho} = \hat{Q} \otimes \hat{M}
\end{equation}
with 
\begin{equation}
\begin{aligned}
\hat{Q} &= Q_{00} \ketbra{0_L} + Q_{11}  \ketbra{1_L} \\
&+ Q_{01} \ketbra{0_L}{1_L} + Q_{10}  \ketbra{1_L}{0_L}
\end{aligned}
\end{equation}
a $2\times2$ matrix defining the  logical qubit, and $\hat{M}$ a generic, possibly mixed, density matrix of a noisy system.
If these conditions are satisfied, we can identify 
\begin{equation}\begin{split} \label{Eq:Minimal_condition_encoding}
    \hat{\rho}_{\pm X} & \equiv \frac{\ketbra{0_L}+\ketbra{1_L} \pm \ketbra{0_L}{1_L}\mp \ketbra{1_L}{0_L}}{2} \otimes \hat{M}, \\
    \hat{\rho}_{\pm Y} & \equiv \frac{\ketbra{0_L}+\ketbra{1_L} \mp i \ketbra{0_L}{1_L}\pm i \ketbra{1_L}{0_L}}{2} \otimes \hat{M}, \\
    \hat{\rho}_{+Z}    & \equiv \ketbra{0_L} \otimes \hat{M},       \quad     \hat{\rho}_{-Z}  \equiv \ketbra{1_L} \otimes \hat{M}.         
    \end{split}
\end{equation}

Operationally, to verify the validity of an encoding we proceed as follows: having identified possible candidates for $\hat{\rho}_{\pm X}$, $\hat{\rho}_{\pm Y}$, and $\hat{\rho}_{\pm Z}$, we verify the existence of a unitary transformation connecting them. To do so, we diagonalize $\hat{\rho}_{+ Z}$ and $\hat{\rho}_{- Z}$ which, being orthogonal, commute, and thus admit a common basis.
By checking that their eigenvalues coincide, we verify that the matrix $\hat M$ associated to these states is the same. We then implement the permutation required for Eq.~\eqref{Eq:Minimal_condition_encoding} to be satisfied, and call the combination of diagonalization and permutation of the two matrices respectively $\hat{R}_{+Z}$ and $\hat{R}_{-Z}$.
Finally, we verify that Eq.~\eqref{Eq:Minimal_condition_encoding} is also satisfied by
\begin{equation}
\label{Eq:Zdiag_rotation}
\begin{split}
&\hat{R}_{+Z} \hat{R}_{-Z} \hat{\rho}_{\pm X} \hat{R}_{-Z}^\dagger \hat{R}_{+Z} ^\dagger,  \\
&\hat{R}_{+Z} \hat{R}_{-Z} \hat{\rho}_{\pm Y} \hat{R}_{-Z}^\dagger \hat{R}_{+Z} ^\dagger, 
\end{split}
\end{equation}
with the same $\hat{M}$.

Note that this construction guarantees only the possibility to encode quantum information and perform gates at a single given time. The possibility of extending the validity of the encoding over time is tied to the action of the Liouvillian on $\operatorname{span}\{\hat{\rho}_{\pm X}, \, \hat{\rho}_{\pm Y}, \, \hat{\rho}_{\pm Z}\}$.
Indeed, in the same way in which we have identified the isomorphism between the density matrices $\hat{\rho}_{\pm X}, \, \hat{\rho}_{\pm Y}, \, \hat{\rho}_{\pm Z}$ and the logical encoding, we can separate the action of the Liouvillian as
\begin{equation}
\LL = \LL_Q \oplus  \LL_M \oplus \LL_{M\to Q} \oplus \LL_{Q\to M}.
\end{equation}
While $\LL_Q$ and $\LL_M$ represent the action of the Liouvillian within $\hat Q$ and $\hat M$, $\LL_{M\to Q} $ and $\LL_{Q\to M}$ describe the processes connecting the logical manifold with the remainder of the Hilbert space containing $\hat{M}$.
The different scenarios identified by Eqs.~\eqref{Eq:general_time_evolution}--\eqref{Eq:metastable_encoding} correspond to the 
different ways in which $\LL $ can act on the code space. Both the steady-state and metastable encodings discussed in Sec.~\ref{sec:encoding} can be seen as an extension of the ideas laid out for the ideal noiseless subsystem structure \cite{lidar1998, fortunato2003, albert2016, lidar2014, kempe2001, knill2000} which we briefly review below in relation to the driven-dissipative Kerr resonator.

\subsection{Noiseless subsystem encoding}
\label{app:encoding_NS}

A noiseless subsystem is formally a subset of the Liouville space immune to the action of dissipation, making the logical information encoded therein transparent to any kind of error process. Indeed, the matrix $\hat Q$ in Eq.~\eqref{Eq:NS} characterizing this encoding is stationary throughout the dynamics [c.f.~Eq.~\eqref{Eq:noiseless_encoding}]. This is always the case whenever $\hat{\rho}_{\pm X}$, $\hat{\rho}_{\pm Y}$, and $\hat{\rho}_{\pm Z}$ span the kernel of the Liouvillian. 
As detailed in Ref.~\cite{lieu2020}, this requirement is equivalent to having four stationary processes: two steady states and two steady coherences. For the strongly $\mathcal{Z}_2$ symmetric Kerr resonator under investigation, this condition is only fulfilled for $\Delta = \kappa_1= \kappa_\phi = 0$, that is, when the only dissipative process is two-photon loss. 

\begin{figure*}[t]
\includegraphics[width=1.\textwidth]{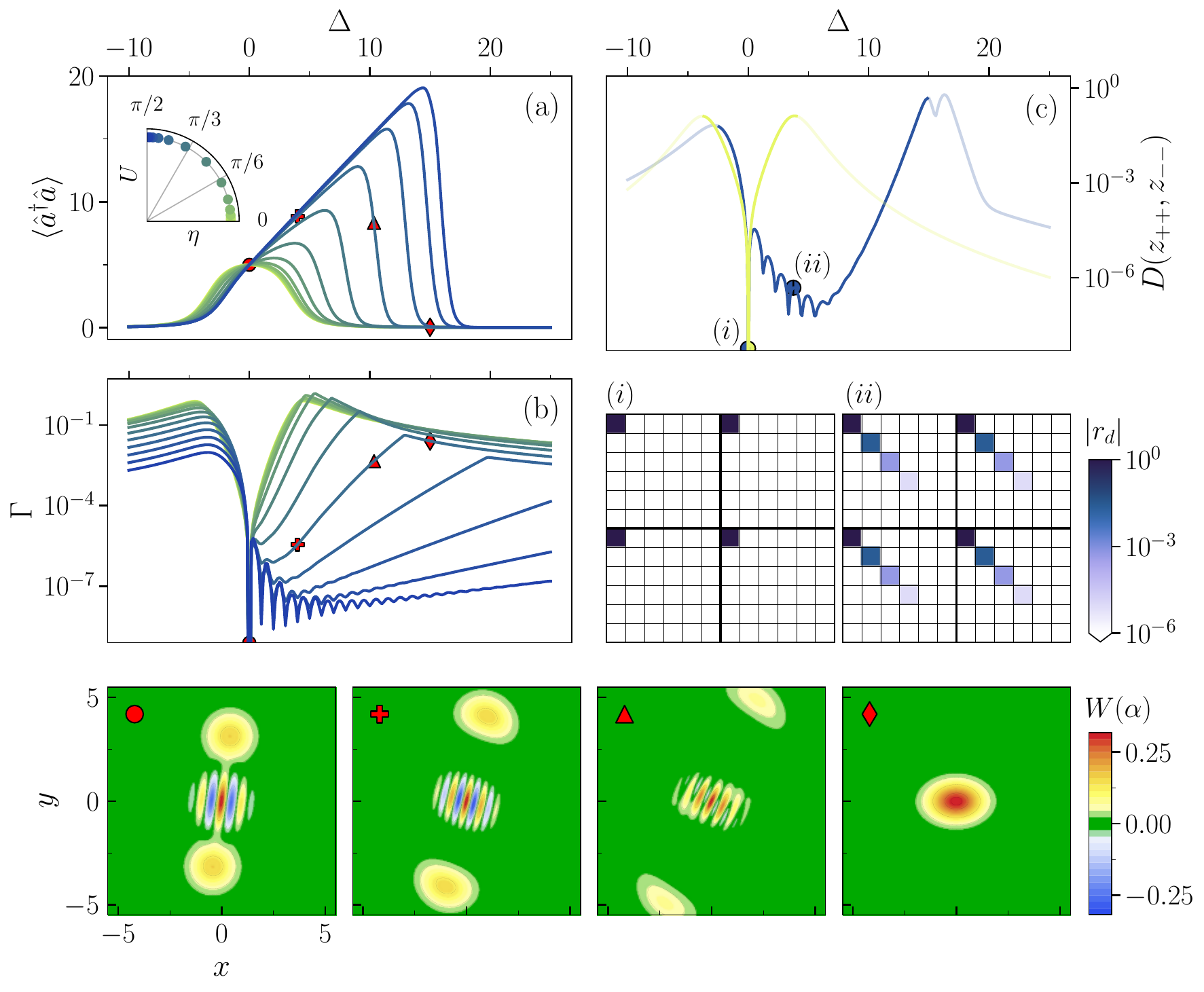} 
\caption{Liouvillian spectral analysis for $G=5$ and different values of $\theta$ in Eq.~\eqref{Eq:W_definition}.
(a) Average occupation of the cavity as a function of the detuning $\Delta$ for different values of the Kerr nonlinearity $U$ and two-photon loss $\eta$.
As $|\alpha(\Delta=0)|^2 = G/W$, all curves intersect at $\Delta=0$. The jump in $\expval*{\hat{a}^\dagger \hat{a}}$ appearing at $\Delta>0$ becomes sharper for larger values of $\theta$, heralding the emergence of a first-order phase transition. (b) Liouvillian gap of the $+-$ symmetry sector $\Lambda_0^{+-}=\Gamma$. Optimal configurations with non-vanishing detuning appear as local minima of the curves and allow taking advantage of the nonlinearity (see text). In addition to this general trend, $\Gamma$ exhibits dips for specific detuning values, namely $\Delta=mU$ with $m\in\mathbb{N}$. These dips become more pronounced as the system approaches the Kerr limit, where they have been recently measured \cite{Venkatraman2022}.
Insets to these plots display the Wigner distribution of $\ket{1_L}$ for $U/\eta = 5$ and increasing detuning values identified by red markers.
(c) In an ideal noiseless subsystem, the eigenoperators $\hat{\rho}_0^{\mu\nu}$ obey $\hat{R}_\mu\rho_{0}^{\mu \nu} \hat{R}_\nu^\dagger \equiv \ketbra{\mu}{\nu}\otimes z_{\mu \nu}$, where $z_{\mu\nu} = z_{\mu^\prime \nu^\prime}$ $\forall \mu,\nu,\mu^\prime,\nu^\prime$ and $\hat{R}_\mu$ is the unitary transformation diagonalizing the density matrix $\hat{\rho}^{\mu\mu}_0$ \cite{lieu2020}. 
We quantify the discrepancy from this ideal case using the trace distance between the two diagonal blocks $z_{++}$ and $z_{--}$. We show this as a function of $\Delta$ for $2\theta/\pi = 0.01$, $0.96$. 
The insets show the structure of the full density matrix for the points indicated in the main figure. $(i)$ exemplifies a decoherence-free subspace, $(ii)$ a noiseless subsystem with a mixed $\hat M$.}
\label{fig:Lspectrum} 
\end{figure*}

Zero eigenvalues of the Liouvillian, however, are not only the hallmark of quantum information encoding, but also one of the main signatures of DPTs. 
Indeed, the four-fold steady-state degeneracy necessary for an exact noiseless subsystem encoding can be achieved at $\Delta\neq 0$ through spontaneous breaking of the underlying symmetry of the model \cite{minganti2018, lieu2020}. While this consideration is exact only in the thermodynamic limit of a DPT, precursors to these effects can be harnessed in \textit{finite-size} systems. The question thus becomes: over which detuning region can an approximate noiseless subsystem encoding exist, and to what extent does the introduction of a nonvanishing detuning destroy the encoding? 

In a Liouvillian framework, the loss of quantum information introduced by the addition of a non-vanishing detuning is determined by the slowest relaxation rate of the off-diagonal Liouvillian sector, i.e. by $\Lambda_0^{+-}=\Lambda_0^{-+}=\Gamma$  [c.f.~Eq.~\eqref{eq:decay_rate}]. Indeed, while the strong $\mathcal{Z}_2$ symmetry ensures $\Lambda_0^{++}=\Lambda_0^{--}=0$, the off-diagonal Liouvillian gap $\Lambda_0^{\pm, \mp}$ can still take finite values and should be minimized in order to suppress qubit decoherence. This quantity is shown in Fig.~\ref{fig:Lspectrum}(b) and allows identifying two regimes.

In the $U=0$ case investigated in Ref.~\cite{lieu2020}, the system displays \emph{two} second-order phase transitions at $\Delta/G = \pm 1$ which symmetrically divide the phase space into a normal region ($\abs{\Delta}/G>1$), hosting at most a classical qubit structure, and a $\mathcal{Z}_2$-broken region ($\abs{\Delta}/G<1$), where $\Lambda_0^{+-}$ is small (zero in the TDL) and an approximate (exact in the TDL) noiseless subsystem structure emerges. For such vanishing nonlinearities the broken symmetry region is very narrow, the maximal photon number is attained at $\Delta = 0$, and can be only increased by increasing $G$. 

For finite $U$, the behaviour of the system drastically changes, as the extent of the $\mathcal{Z}_2$-broken region wherein $\Lambda_0^{+-}\simeq 0$ increases far beyond the typical values obtained for $U \simeq 0$.  
Indeed, while the phase boundary for $\Delta<0$ is not significantly modified, and the corresponding DPT is still of second order, for $\Delta>0$ a \textit{first-order} transition emerges at $\Delta/G \simeq \sqrt{1 + (U/\eta)^2}$ replacing the second order one at $\Delta/G=1$ for $U=0$. This allows for an approximate noiseless subsystem tensor structure to hold over a much broader region of positive detuning. 
In this regard, the use of a first-order DPT entails several advantages with respect to the proposal of Ref.~\cite{lieu2020}.

Identifying the gradual transition from a second to a first-order DPT are the elbow-like shape of the photon number for $U=0$ in Fig.~\ref{fig:Lspectrum}(a), gradually evolving into a jump discontinuity for increasing values of $U$, and the precursors to a point-like closure \cite{minganti2018} of the diagonal gap $\Lambda_1^{++}$ seen in Fig.~\ref{fig:diagonalgap}(a) of the main text.

Numerical evidence of the approximate fulfillment of Eq.~\eqref{Eq:Minimal_condition_encoding} is provided in Fig.~\ref{fig:Lspectrum}(c). As explained above and in Ref.~\cite{lieu2020}, we quantify the discrepancy from the ideal case via the trace distance $D(A,B) = \sqrt{\operatorname{Tr}\left\{(A-B)^\dagger (A-B)\right\}}$ between the diagonal forms of $\hat\rho_{+Z}$ and $\hat\rho_{-Z}$.
While in the dissipative limit $D(\hat\rho_{+Z}, \hat\rho_{-Z})$ only approaches zero in the vicinity of $\Delta=0$, as the ratio of $U/\eta$ is increased, small values of $D(\hat\rho_{+Z}, \hat\rho_{-Z})$ persist over a much wider range of positive detunings. Similar results were found for the distance from the orthogonal axes rotated according to Eq.~\eqref{Eq:Zdiag_rotation} (not shown). The insets below Fig.~\ref{fig:Lspectrum}(c) are color plots of the absolute value of the matrix elements, for the two cases highlighted in the plot. They illustrate, respectively, a decoherence-free subspace steady state spanned by pure cat states, and a noiseless subsystem steady state spanned by mixed states.

We conclude that a \textit{steady} cat-like state, capable of encoding quantum information, can be generated in the whole region $-G < \Delta < G \sqrt{1 + (U/\eta)^2}$. 
This state -- whose Wigner representation is shown in the insets to Figs.~\ref{fig:Lspectrum}(a-b) for increasing values of $\Delta$ -- is not exactly the ideal cat described in Eq.~\eqref{Eq:definition_cat}, though it displays analogous properties.

\subsection{Steady-state and metastable encoding}

Having reviewed the ideal noiseless subsystem encoding, we generalize it to include biased-noise encoding and the possibility to encode the system on metastable states.

The full Liouvillian $\LL$ in Eq.~\eqref{eq:full_cat} only possesses a weak $\mathcal{Z}_2$ symmetry, its steady state $\sss = \hat\rho_0^+$ is unique, and its generic density matrix decomposes along its eigenoperators as
\begin{equation}\label{Eq:decomposition}
\hat\rho(t) = \sss + \sum_{j\geq0} c_j^-(t) \hat\rho_j^- +  \sum_{j>0} c_j^+(t) \hat\rho_j^+.
\end{equation}
Note that $\hat\rho_j^\mu$ are not physical density matrices as they are in general neither positive nor have unit trace \footnote{This is true for all $\hat\rho_j^\mu$ except for $\hat\rho_0^+ \equiv \sss$. To simplify notation, we will refer to $\hat\rho_j^\mu$ as any element in $\{\hat\rho_j^\mu\}_{\mu\in\{\pm\}, j\in\mathbb{N}}/\{\hat\rho_0^+\}$ and use $\sss$ to identify $\hat\rho_0^+$. 
},
and they acquire a physical meaning only through their eigendecomposition described in Eq.~\eqref{Eq:diagonalizaion_DM}.
Following the main text, we distinguish two regimes of operation wherein the nature of the encoding drastically differs. 

\begin{figure}[ht]
\includegraphics[width=0.48\textwidth]{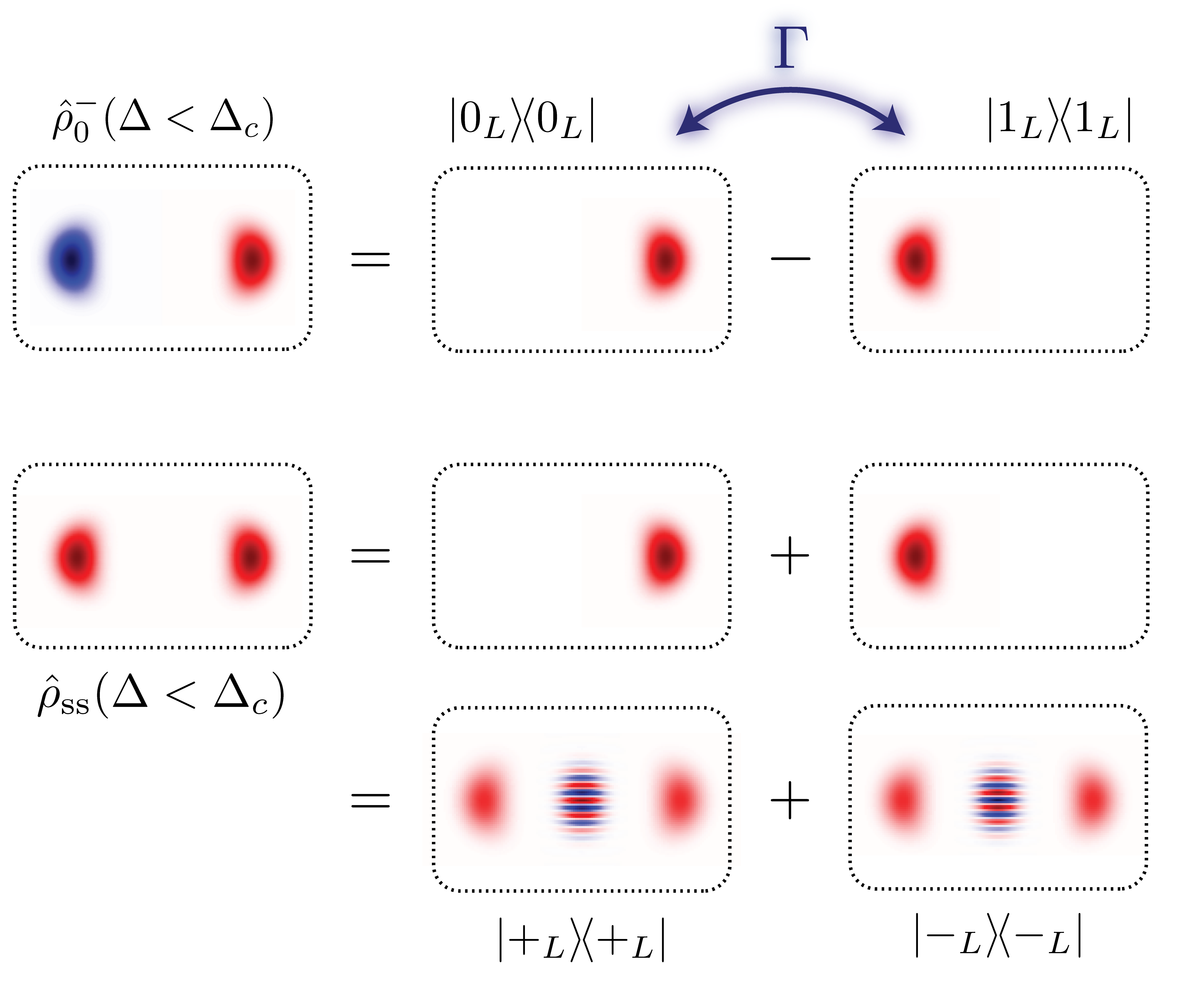} 
\caption{\label{fig:SS_schematic} Schematic representation of the states comprising the steady-state encoding defined in Eq.~\ref{eqn:encoding_SS}. We show $\hat\rho_{\pm Z}$ and $\hat \rho_{\pm X}$. $\hat \rho_{\pm Y}$ can be obtained by acting on $\hat\rho_{\pm Z}$ with the rotation $\hat X_{\varphi} = \cos\frac{\varphi}{2} (\hat \rho_{+ Z} + \hat \rho_{- Z}) + i\sin\frac{\varphi}{2} (\hat \rho_{+ X} - \hat \rho_{- X})$ with $\varphi = \pi/2$.}
\end{figure}

Dropping the $\otimes \hat{M}$ notation for brevity, for $\Delta<\Delta_c$, the logical $Z$ and $X$-axis are identified by
\begin{equation}
\label{eqn:encoding_SS}
\begin{aligned}
    \hat\rho_0^- &= \hat\rho_0^{-}(+) - \hat\rho_0^{-}(-)\equiv \ketbra{0_L} -\ketbra{1_L}\\[0.2cm]
    \sss &= \sss(+) + \sss(-) = \ketbra{+_L} + \ketbra{-_L} \\
     & = \ketbra{0_L} +\ketbra{1_L}.
\end{aligned}
\end{equation}
In this regime, the unique steady state towards which the system decays coincides with the center of the logical Bloch sphere so that, although the matrix $\hat Q(t)$ evolves in time [c.f.~Eq.~\eqref{Eq:statedy-state_encoding}], the dynamics is restricted to the code manifold. Specifically, the only process acting on the poles of the logical Bloch sphere is one which exchanges them at a rate $\Gamma=\Lambda_0^-$ (bit-flip error).  
The schematic representation of this encoding, which we deem \textit{steady-state encoding}, is shown in Fig.~\ref{fig:SS_schematic}.

\begin{figure}[ht]
\includegraphics[width=0.48\textwidth]{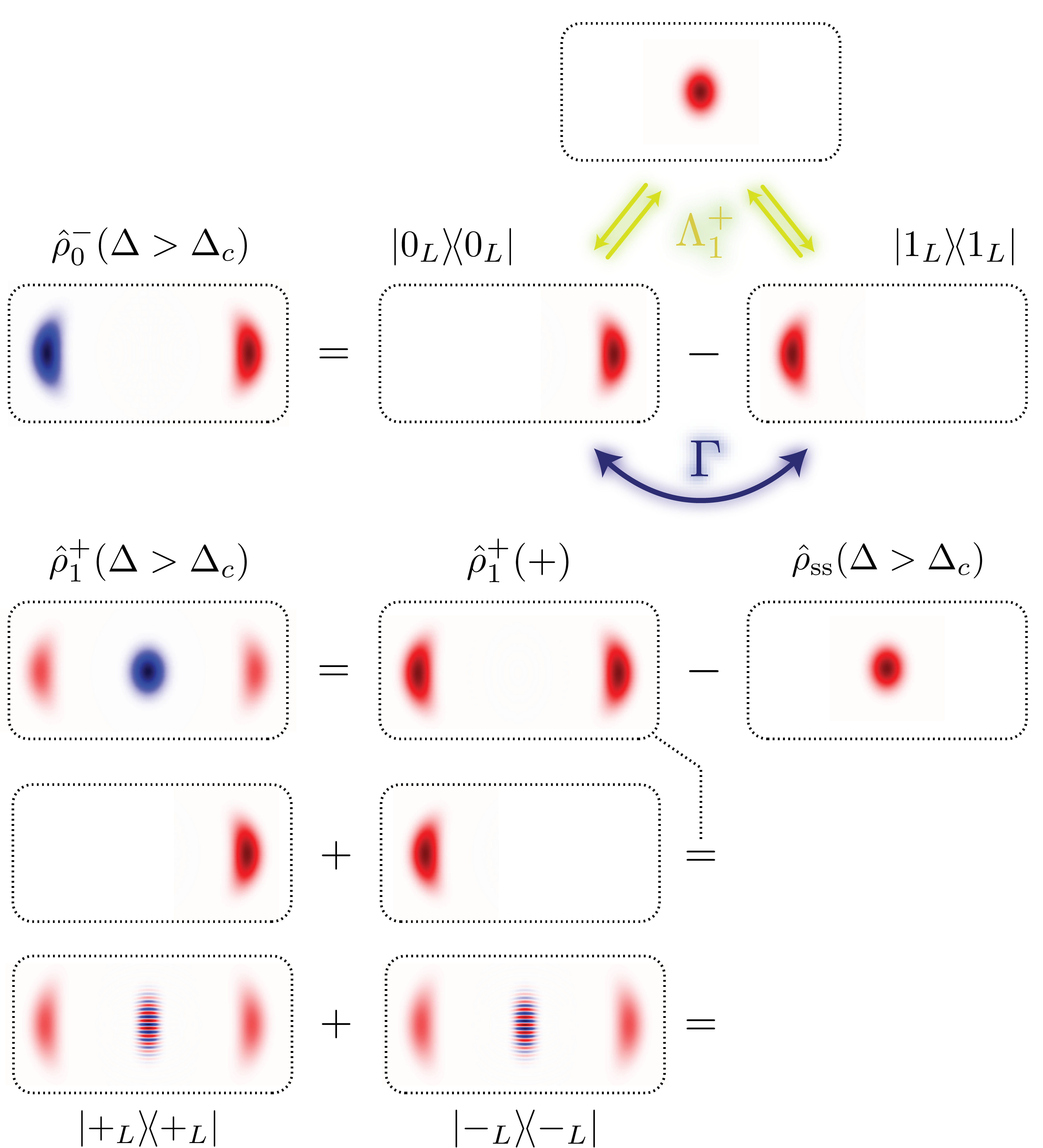} 
\caption{\label{fig:MS_schematic} Schematic representation of the states comprising the steady-state encoding defined in Eq.~\ref{eqn:encoding_MS}.}
\end{figure}

For $\Delta>\Delta_c$, instead, the steady state approximately coincides with the vacuum and is no longer a viable choice for encoding the logical $X$-axis. 
Nevertheless, the emergence of a second (almost) infinitely long-lived process ($\hat\rho_1^+$) spanned by highly populated states still allows identifying the logical $Z$ and $X$-axis as
\begin{equation}
\label{eqn:encoding_MS}
\begin{aligned}
    &\hat\rho_0^- = \hat\rho_0^{-}(+) - \hat\rho_0^{-}(-)\equiv \ketbra{0_L} -\ketbra{1_L}\\[0.2cm]
    &\hat\rho_1^+(+) = \hat\rho_1^+(++) + \hat\rho_1^+(+-) = \ketbra{+_L} + \ketbra{-_L}.
\end{aligned}
\end{equation}
The rates at which these states decay can be obtained using the eigendecomposition in Eq.~\eqref{Eq:decomposition}.
What we find is that, as depicted in Fig.~\ref{fig:MS_schematic}, \textit{only one} additional channel emerges which connects the poles of the Bloch sphere with the vacuum at a rate $\Lambda_1^+$, whose dependence on $\Delta$ has been characterized in Figs.~\ref{fig:deltascan}~and~\ref{fig:diagonalgap} of the main text. 
The metastable state $\rho_1^+(+)$ replaces the steady state as the center of the logical Bloch sphere allowing us to extend the construction of steady-state encoding to incorporate metastability in what we dub a \textit{metastable encoding}, whereby Eq.~\eqref{Eq:metastable_encoding} follows.

\begin{figure}[ht]
\includegraphics[width=0.48\textwidth]{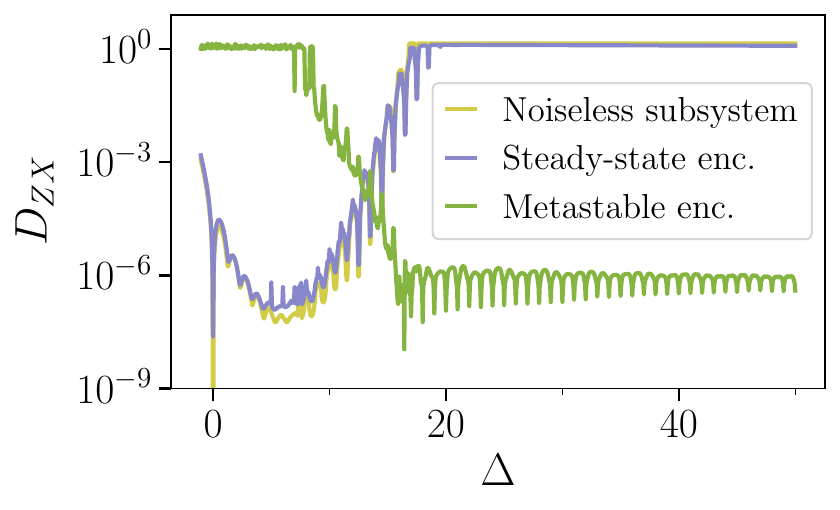} 
\caption{\label{fig:MS_dist} Same distance quantifier as that used in Fig.~\ref{fig:Lspectrum}(c). Here it evaluated between $\hat\rho_Z$ and $\hat \rho_X$ following the procedure detailed in the main text.  We plot the results as a function of the detuning $\Delta$ for $\kappa_1 = 10^{-3}\gg \kappa_\phi= 10^{-5}$ and $G=5$. The dips at integer values of detuning are to be imputed to the tunnelling suppression effects of Hamiltonian nature discussed in Ref.~\cite{Venkatraman2022}.}
\end{figure}

Finally, in Fig.~\ref{fig:MS_dist} we provide numerical evidence for the qubit structures defined in Eqs.~\eqref{eqn:encoding_SS}~and~\eqref{eqn:encoding_MS} to  satisfy Eq.~\eqref{Eq:NS}. Once again we do so by evaluating the trace distance between the diagonal form of $\hat\rho_{\pm Z}$ and the matrix obtained by applying to $\hat\rho_{+X}$ the transformation in Eq.~\eqref{Eq:Zdiag_rotation}.
As $\Delta$ is increased, $U/\eta$ is modified accordingly from point to point to ensure that $\Delta=\operatorname{min}[\Delta_{\rm opt}(\theta),\Delta_{\rm max}]$. We can clearly distinguish two regions: $\Delta<\Delta_c$ where the steady-state encoding approximates Eq.~\ref{Eq:NS} and the metastable one completely fails, and  $\Delta>\Delta_c$ where the converse is true.

\section{Extended analysis of $\Gamma$ and $\Gamma_{\rm leak}$}
\label{sec:app_extension}
\begin{figure*}[t]
\includegraphics[width=\textwidth]{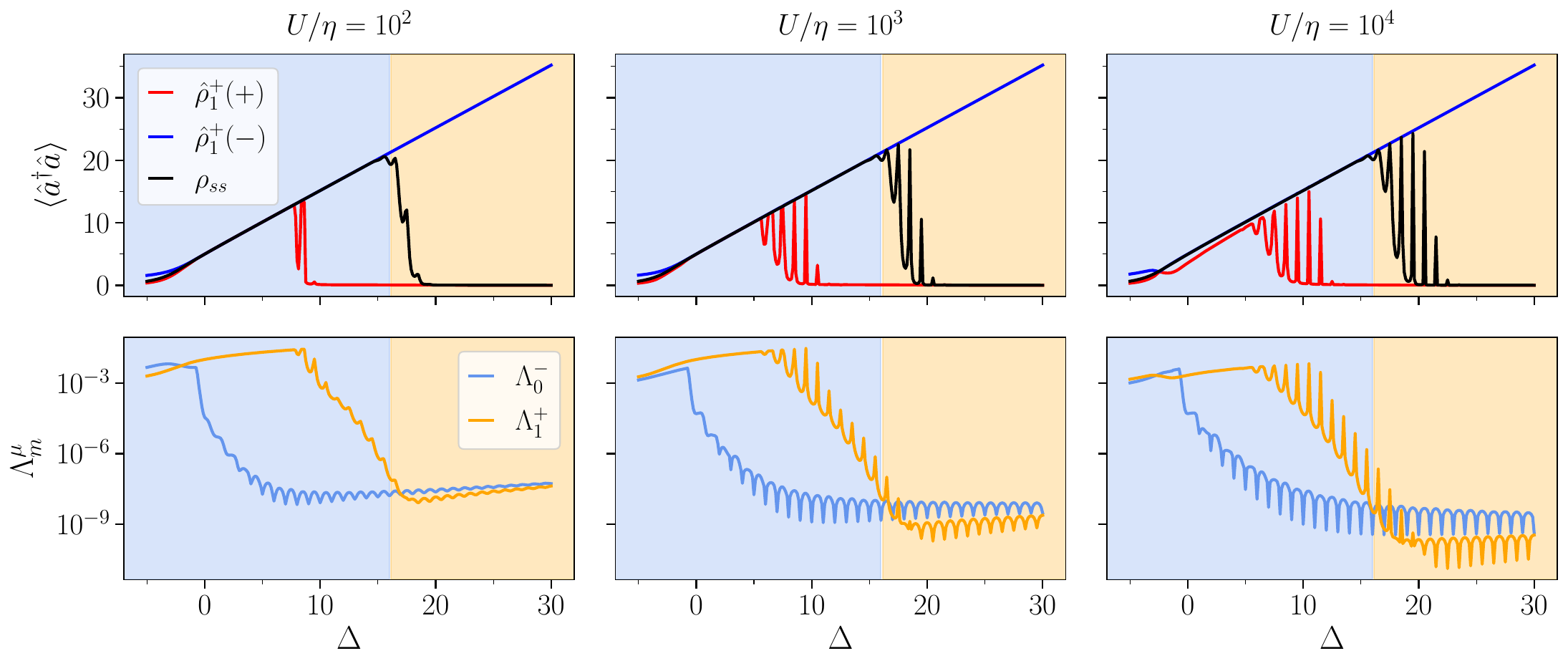} 
\caption{\label{fig:deltascan_additional_pldominated} The average photon number of the steady and metastable states as defined in the main text, and the Liouvillian gaps $\Lambda_1^+$ and $\Lambda_0^-$ as defined in Eq.~\eqref{Eq:block_eigenvalues_weak}, are shown as a function of detuning $\Delta$. The shaded blue (orange) area identifies the region hosting a steady-state (metastable) encoding. We consider $G=5$ and $\kappa_1 = 10^{-3}\gg \kappa_\phi= 10^{-5}$.}
\end{figure*}
\begin{figure*}[t]
\includegraphics[width=\textwidth]{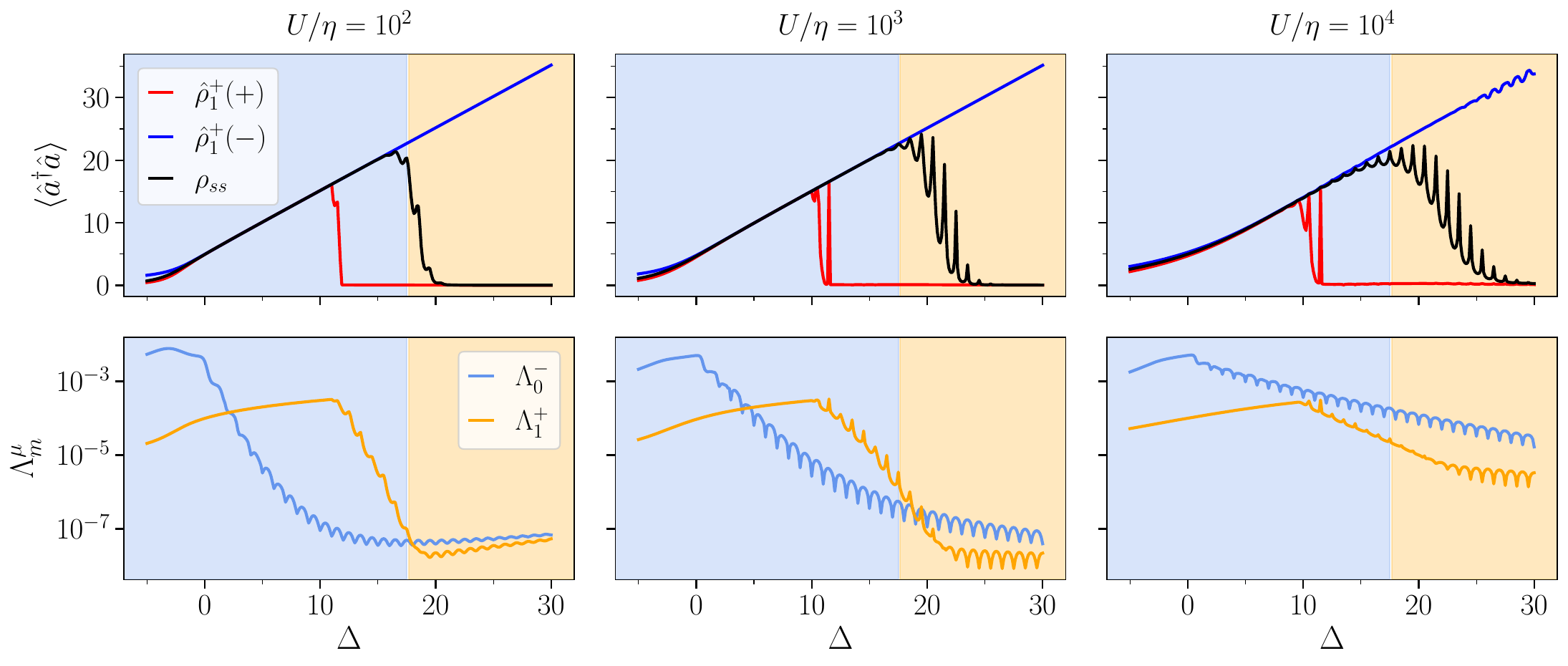} 
\caption{\label{fig:deltascan_additional_dephdominated} Same as Fig.~\ref{fig:deltascan_additional_pldominated} but for $\kappa_1 = 10^{-5}\ll \kappa_\phi= 10^{-3}$.}
\end{figure*}
We extend the discussion on the relation between $\Gamma$  and $\Gamma_{\rm leak}$ started in Sec.~\ref{sec:encoding} by repeating the same analysis presented in Figs.~\ref{fig:deltascan}(a)~and~(b) for different choices of $U/\eta$, and for both noise configurations. We observe that the peak-like structure characterizing both $\Lambda_0^-$ and $\Lambda_1^+$ discussed in the main text is suppressed when increasing $\eta$. Moreover, as expected from Figs.~\ref{fig:gamma_v_Delta_v_U}(a)~and~(b), we note that the optimal detuning value minimizing $\Lambda_0^-$ tends towards the steady-state encoding region (blue area) the higher the value of $\eta$ (lower ratios $U/\eta$) one considers. This encoding is not affected by any metastable timescale, thus voiding the tradeoff between $\Lambda_0^-$ and $\Lambda_1^+$ identified in the metastable encoding (orange area).


\begin{thebibliography}{66}%
	\makeatletter
	\providecommand \@ifxundefined [1]{%
		\@ifx{#1\undefined}
	}%
	\providecommand \@ifnum [1]{%
		\ifnum #1\expandafter \@firstoftwo
		\else \expandafter \@secondoftwo
		\fi
	}%
	\providecommand \@ifx [1]{%
		\ifx #1\expandafter \@firstoftwo
		\else \expandafter \@secondoftwo
		\fi
	}%
	\providecommand \natexlab [1]{#1}%
	\providecommand \enquote  [1]{``#1''}%
	\providecommand \bibnamefont  [1]{#1}%
	\providecommand \bibfnamefont [1]{#1}%
	\providecommand \citenamefont [1]{#1}%
	\providecommand \href@noop [0]{\@secondoftwo}%
	\providecommand \href [0]{\begingroup \@sanitize@url \@href}%
	\providecommand \@href[1]{\@@startlink{#1}\@@href}%
	\providecommand \@@href[1]{\endgroup#1\@@endlink}%
	\providecommand \@sanitize@url [0]{\catcode `\\12\catcode `\$12\catcode
		`\&12\catcode `\#12\catcode `\^12\catcode `\_12\catcode `\%12\relax}%
	\providecommand \@@startlink[1]{}%
	\providecommand \@@endlink[0]{}%
	\providecommand \url  [0]{\begingroup\@sanitize@url \@url }%
	\providecommand \@url [1]{\endgroup\@href {#1}{\urlprefix }}%
	\providecommand \urlprefix  [0]{URL }%
	\providecommand \Eprint [0]{\href }%
	\providecommand \doibase [0]{http://dx.doi.org/}%
	\providecommand \selectlanguage [0]{\@gobble}%
	\providecommand \bibinfo  [0]{\@secondoftwo}%
	\providecommand \bibfield  [0]{\@secondoftwo}%
	\providecommand \translation [1]{[#1]}%
	\providecommand \BibitemOpen [0]{}%
	\providecommand \bibitemStop [0]{}%
	\providecommand \bibitemNoStop [0]{.\EOS\space}%
	\providecommand \EOS [0]{\spacefactor3000\relax}%
	\providecommand \BibitemShut  [1]{\csname bibitem#1\endcsname}%
	\let\auto@bib@innerbib\@empty
	\bibitem [{\citenamefont {Lidar}\ and\ \citenamefont {Brun}(2013)}]{lidar2013}%
	\BibitemOpen
	\bibinfo {editor} {D.~A. Lidar}\ and\ \bibinfo {editor} {T.~A. Brun},\ eds.,\
	\href {\doibase10.1017/CBO9781139034807} {\emph {\bibinfo {title} {Quantum
				{{Error Correction}}}}}\ (\bibinfo  {publisher} {{Cambridge University
			Press}},\ \bibinfo {address} {{Cambridge}},\ \bibinfo {year}
	{2013})\BibitemShut {NoStop}%
	\bibitem [{\citenamefont {Haroche}\ and\ \citenamefont
		{Raimond}(2013)}]{haroche2013}%
	\BibitemOpen
	\bibinfo {author} {S.~Haroche}\ and\ \bibinfo {author} {J.-M. Raimond},\
	\href@noop {} {\emph {\bibinfo {title} {Exploring the Quantum: Atoms,
				Cavities, and Photons}}}\ (\bibinfo  {publisher} {{Oxford University
			Press}},\ \bibinfo {address} {{Oxford}},\ \bibinfo {year} {2013})\BibitemShut
	{NoStop}%
	\bibitem [{\citenamefont {Preskill}(2018)}]{preskill2018}%
	\BibitemOpen
	\bibinfo {author} {J.~Preskill},\ \emph {\bibinfo {title} {Quantum
			{{Computing}} in the {{NISQ}} Era and Beyond}},\ \href
	{\doibase10.22331/q-2018-08-06-79} {\bibfield  {journal} {\bibinfo  {journal}
			{Quantum}\ }\textbf {\bibinfo {volume} {2}},\ \bibinfo {pages} {79} (\bibinfo
		{year} {2018})}\BibitemShut {NoStop}%
	\bibitem [{\citenamefont {Breuer}\ and\ \citenamefont
		{Petruccione}(2007)}]{breuer2007}%
	\BibitemOpen
	\bibinfo {author} {H.-P. Breuer}\ and\ \bibinfo {author} {F.~Petruccione},\
	\href {\doibase10.1093/acprof:oso/9780199213900.001.0001} {\emph {\bibinfo
			{title} {The {{Theory}} of {{Open Quantum Systems}}}}}\ (\bibinfo
	{publisher} {{Oxford University Press}},\ \bibinfo {address} {{Oxford}},\
	\bibinfo {year} {2007})\BibitemShut {NoStop}%
	\bibitem [{\citenamefont {Nielsen}\ and\ \citenamefont
		{Chuang}(2011)}]{nielsen2011}%
	\BibitemOpen
	\bibinfo {author} {M.~A. Nielsen}\ and\ \bibinfo {author} {I.~L. Chuang},\
	\href@noop {} {\emph {\bibinfo {title} {Quantum {{Computation}} and {{Quantum
						Information}}: 10th {{Anniversary Edition}}}}},\ \bibinfo {edition} {tenth}\
	ed.\ (\bibinfo  {publisher} {{Cambridge University Press}},\ \bibinfo
	{address} {{USA}},\ \bibinfo {year} {2011})\BibitemShut {NoStop}%
	\bibitem [{\citenamefont {Campbell}\ \emph {et~al.}(2017)\citenamefont
		{Campbell}, \citenamefont {Terhal},\ and\ \citenamefont
		{Vuillot}}]{campbell2017}%
	\BibitemOpen
	\bibinfo {author} {E.~T. Campbell}, \bibinfo {author} {B.~M. Terhal}\ and\
	\bibinfo {author} {C.~Vuillot},\ \emph {\bibinfo {title} {Roads towards
			Fault-Tolerant Universal Quantum Computation}},\ \href
	{\doibase10.1038/nature23460} {\bibfield  {journal} {\bibinfo  {journal}
			{Nature}\ }\textbf {\bibinfo {volume} {549}},\ \bibinfo {pages} {172}
		(\bibinfo {year} {2017})}\BibitemShut {NoStop}%
	\bibitem [{\citenamefont {Terhal}(2015)}]{terhal2015}%
	\BibitemOpen
	\bibinfo {author} {B.~M. Terhal},\ \emph {\bibinfo {title} {Quantum Error
			Correction for Quantum Memories}},\ \href {\doibase10.1103/RevModPhys.87.307}
	{\bibfield  {journal} {\bibinfo  {journal} {Reviews of Modern Physics}\
		}\textbf {\bibinfo {volume} {87}},\ \bibinfo {pages} {307} (\bibinfo {year}
		{2015})}\BibitemShut {NoStop}%
	\bibitem [{\citenamefont {Gottesman}\ \emph {et~al.}(2001)\citenamefont
		{Gottesman}, \citenamefont {Kitaev},\ and\ \citenamefont
		{Preskill}}]{gottesman2001}%
	\BibitemOpen
	\bibinfo {author} {D.~Gottesman}, \bibinfo {author} {A.~Kitaev}\ and\
	\bibinfo {author} {J.~Preskill},\ \emph {\bibinfo {title} {Encoding a Qubit
			in an Oscillator}},\ \href {\doibase10.1103/PhysRevA.64.012310} {\bibfield
		{journal} {\bibinfo  {journal} {Physical Review A}\ }\textbf {\bibinfo
			{volume} {64}},\ \bibinfo {pages} {012310} (\bibinfo {year}
		{2001})}\BibitemShut {NoStop}%
	\bibitem [{\citenamefont {Mirrahimi}\ \emph {et~al.}(2014)\citenamefont
		{Mirrahimi}, \citenamefont {Leghtas}, \citenamefont {Albert}, \citenamefont
		{Touzard}, \citenamefont {Schoelkopf}, \citenamefont {Jiang},\ and\
		\citenamefont {Devoret}}]{mirrahimi2014}%
	\BibitemOpen
	\bibinfo {author} {M.~Mirrahimi}, \bibinfo {author} {Z.~Leghtas}, \bibinfo
	{author} {V.~V. Albert}, \bibinfo {author} {S.~Touzard}, \bibinfo {author}
	{R.~J. Schoelkopf}, \bibinfo {author} {L.~Jiang}\ and\ \bibinfo {author}
	{M.~H. Devoret},\ \emph {\bibinfo {title} {Dynamically Protected Cat-Qubits:
			A New Paradigm for Universal Quantum Computation}},\ \href
	{\doibase10.1088/1367-2630/16/4/045014} {\bibfield  {journal} {\bibinfo
			{journal} {New Journal of Physics}\ }\textbf {\bibinfo {volume} {16}},\
		\bibinfo {pages} {045014} (\bibinfo {year} {2014})}\BibitemShut {NoStop}%
	\bibitem [{\citenamefont {Cai}\ \emph {et~al.}(2021)\citenamefont {Cai},
		\citenamefont {Ma}, \citenamefont {Wang}, \citenamefont {Zou},\ and\
		\citenamefont {Sun}}]{cai2021}%
	\BibitemOpen
	\bibinfo {author} {W.~Cai}, \bibinfo {author} {Y.~Ma}, \bibinfo {author}
	{W.~Wang}, \bibinfo {author} {C.-L. Zou}\ and\ \bibinfo {author} {L.~Sun},\
	\emph {\bibinfo {title} {Bosonic Quantum Error Correction Codes in
			Superconducting Quantum Circuits}},\ \href
	{\doibase10.1016/j.fmre.2020.12.006} {\bibfield  {journal} {\bibinfo
			{journal} {Fundamental Research}\ }\textbf {\bibinfo {volume} {1}},\ \bibinfo
		{pages} {50} (\bibinfo {year} {2021})}\BibitemShut {NoStop}%
	\bibitem [{\citenamefont {Joshi}\ \emph {et~al.}(2021)\citenamefont {Joshi},
		\citenamefont {Noh},\ and\ \citenamefont {Gao}}]{joshi2021}%
	\BibitemOpen
	\bibinfo {author} {A.~Joshi}, \bibinfo {author} {K.~Noh}\ and\ \bibinfo
	{author} {Y.~Y. Gao},\ \emph {\bibinfo {title} {Quantum Information
			Processing with Bosonic Qubits in Circuit {{QED}}}},\ \href
	{\doibase10.1088/2058-9565/abe989} {\bibfield  {journal} {\bibinfo  {journal}
			{Quantum Science and Technology}\ }\textbf {\bibinfo {volume} {6}},\ \bibinfo
		{pages} {033001} (\bibinfo {year} {2021})}\BibitemShut {NoStop}%
	\bibitem [{\citenamefont {Terhal}\ \emph {et~al.}(2020)\citenamefont {Terhal},
		\citenamefont {Conrad},\ and\ \citenamefont {Vuillot}}]{terhal2020}%
	\BibitemOpen
	\bibinfo {author} {B.~M. Terhal}, \bibinfo {author} {J.~Conrad}\ and\
	\bibinfo {author} {C.~Vuillot},\ \emph {\bibinfo {title} {Towards Scalable
			Bosonic Quantum Error Correction}},\ \href {\doibase10.1088/2058-9565/ab98a5}
	{\bibfield  {journal} {\bibinfo  {journal} {Quantum Science and Technology}\
		}\textbf {\bibinfo {volume} {5}},\ \bibinfo {pages} {043001} (\bibinfo {year}
		{2020})}\BibitemShut {NoStop}%
	\bibitem [{\citenamefont {Knill}\ \emph {et~al.}(2000)\citenamefont {Knill},
		\citenamefont {Laflamme},\ and\ \citenamefont {Viola}}]{knill2000}%
	\BibitemOpen
	\bibinfo {author} {E.~Knill}, \bibinfo {author} {R.~Laflamme}\ and\ \bibinfo
	{author} {L.~Viola},\ \emph {\bibinfo {title} {Theory of {{Quantum Error
					Correction}} for {{General Noise}}}},\ \href
	{\doibase10.1103/PhysRevLett.84.2525} {\bibfield  {journal} {\bibinfo
			{journal} {Physical Review Letters}\ }\textbf {\bibinfo {volume} {84}},\
		\bibinfo {pages} {2525} (\bibinfo {year} {2000})}\BibitemShut {NoStop}%
	\bibitem [{\citenamefont {Michael}\ \emph {et~al.}(2016)\citenamefont
		{Michael}, \citenamefont {Silveri}, \citenamefont {Brierley}, \citenamefont
		{Albert}, \citenamefont {Salmilehto}, \citenamefont {Jiang},\ and\
		\citenamefont {Girvin}}]{michael2016}%
	\BibitemOpen
	\bibinfo {author} {M.~H. Michael}, \bibinfo {author} {M.~Silveri}, \bibinfo
	{author} {R.~T. Brierley}, \bibinfo {author} {V.~V. Albert}, \bibinfo
	{author} {J.~Salmilehto}, \bibinfo {author} {L.~Jiang}\ and\ \bibinfo
	{author} {S.~M. Girvin},\ \emph {\bibinfo {title} {New {{Class}} of {{Quantum
					Error-Correcting Codes}} for a {{Bosonic Mode}}}},\ \href
	{\doibase10.1103/PhysRevX.6.031006} {\bibfield  {journal} {\bibinfo
			{journal} {Physical Review X}\ }\textbf {\bibinfo {volume} {6}},\ \bibinfo
		{pages} {031006} (\bibinfo {year} {2016})}\BibitemShut {NoStop}%
	\bibitem [{\citenamefont {Albert}(2018)}]{albert2018}%
	\BibitemOpen
	\bibinfo {author} {V.~V. Albert},\ \href@noop {} {\emph {\bibinfo {title}
			{Lindbladians with Multiple Steady States: Theory and Applications}}}
	(\bibinfo {year} {2018}),\ \Eprint {http://arxiv.org/abs/1802.00010}
	{arXiv:1802.00010} \BibitemShut {NoStop}%
	\bibitem [{\citenamefont {Gilles}\ \emph {et~al.}(1994)\citenamefont {Gilles},
		\citenamefont {Garraway},\ and\ \citenamefont {Knight}}]{gilles1994}%
	\BibitemOpen
	\bibinfo {author} {L.~Gilles}, \bibinfo {author} {B.~M. Garraway}\ and\
	\bibinfo {author} {P.~L. Knight},\ \emph {\bibinfo {title} {Generation of
			Nonclassical Light by Dissipative Two-Photon Processes}},\ \href
	{\doibase10.1103/PhysRevA.49.2785} {\bibfield  {journal} {\bibinfo  {journal}
			{Physical Review A}\ }\textbf {\bibinfo {volume} {49}},\ \bibinfo {pages}
		{2785} (\bibinfo {year} {1994})}\BibitemShut {NoStop}%
	\bibitem [{\citenamefont {Leghtas}\ \emph {et~al.}(2015)\citenamefont
		{Leghtas}, \citenamefont {Touzard}, \citenamefont {Pop}, \citenamefont {Kou},
		\citenamefont {Vlastakis}, \citenamefont {Petrenko}, \citenamefont {Sliwa},
		\citenamefont {Narla}, \citenamefont {Shankar}, \citenamefont {Hatridge},
		\citenamefont {Reagor}, \citenamefont {Frunzio}, \citenamefont {Schoelkopf},
		\citenamefont {Mirrahimi},\ and\ \citenamefont {Devoret}}]{leghtas2015}%
	\BibitemOpen
	\bibinfo {author} {Z.~Leghtas}, \bibinfo {author} {S.~Touzard}, \bibinfo
	{author} {I.~M. Pop}, \bibinfo {author} {A.~Kou}, \bibinfo {author}
	{B.~Vlastakis}, \bibinfo {author} {A.~Petrenko}, \bibinfo {author} {K.~M.
		Sliwa}, \bibinfo {author} {A.~Narla}, \bibinfo {author} {S.~Shankar},
	\bibinfo {author} {M.~J. Hatridge}, \bibinfo {author} {M.~Reagor}, \bibinfo
	{author} {L.~Frunzio}, \bibinfo {author} {R.~J. Schoelkopf}, \bibinfo
	{author} {M.~Mirrahimi}\ and\ \bibinfo {author} {M.~H. Devoret},\ \emph
	{\bibinfo {title} {Confining the State of Light to a Quantum Manifold by
			Engineered Two-Photon Loss}},\ \href {\doibase10.1126/science.aaa2085}
	{\bibfield  {journal} {\bibinfo  {journal} {Science}\ }\textbf {\bibinfo
			{volume} {347}},\ \bibinfo {pages} {853} (\bibinfo {year}
		{2015})}\BibitemShut {NoStop}%
	\bibitem [{\citenamefont {Touzard}\ \emph {et~al.}(2018)\citenamefont
		{Touzard}, \citenamefont {Grimm}, \citenamefont {Leghtas}, \citenamefont
		{Mundhada}, \citenamefont {Reinhold}, \citenamefont {Axline}, \citenamefont
		{Reagor}, \citenamefont {Chou}, \citenamefont {Blumoff}, \citenamefont
		{Sliwa}, \citenamefont {Shankar}, \citenamefont {Frunzio}, \citenamefont
		{Schoelkopf}, \citenamefont {Mirrahimi},\ and\ \citenamefont
		{Devoret}}]{touzard2018}%
	\BibitemOpen
	\bibinfo {author} {S.~Touzard}, \bibinfo {author} {A.~Grimm}, \bibinfo
	{author} {Z.~Leghtas}, \bibinfo {author} {S.~O. Mundhada}, \bibinfo {author}
	{P.~Reinhold}, \bibinfo {author} {C.~Axline}, \bibinfo {author} {M.~Reagor},
	\bibinfo {author} {K.~Chou}, \bibinfo {author} {J.~Blumoff}, \bibinfo
	{author} {K.~M. Sliwa}, \bibinfo {author} {S.~Shankar}, \bibinfo {author}
	{L.~Frunzio}, \bibinfo {author} {R.~J. Schoelkopf}, \bibinfo {author}
	{M.~Mirrahimi}\ and\ \bibinfo {author} {M.~H. Devoret},\ \emph {\bibinfo
		{title} {Coherent {{Oscillations}} inside a {{Quantum Manifold Stabilized}}
			by {{Dissipation}}}},\ \href {\doibase10.1103/PhysRevX.8.021005} {\bibfield
		{journal} {\bibinfo  {journal} {Physical Review X}\ }\textbf {\bibinfo
			{volume} {8}},\ \bibinfo {pages} {021005} (\bibinfo {year}
		{2018})}\BibitemShut {NoStop}%
	\bibitem [{\citenamefont {Xu}\ \emph {et~al.}(2022)\citenamefont {Xu},
		\citenamefont {Iverson}, \citenamefont {Brand{\~a}o},\ and\ \citenamefont
		{Jiang}}]{xu2022}%
	\BibitemOpen
	\bibinfo {author} {Q.~Xu}, \bibinfo {author} {J.~K. Iverson}, \bibinfo
	{author} {F.~G. S.~L. Brand{\~a}o}\ and\ \bibinfo {author} {L.~Jiang},\ \emph
	{\bibinfo {title} {Engineering Fast Bias-Preserving Gates on Stabilized Cat
			Qubits}},\ \href {\doibase10.1103/PhysRevResearch.4.013082} {\bibfield
		{journal} {\bibinfo  {journal} {Physical Review Research}\ }\textbf {\bibinfo
			{volume} {4}},\ \bibinfo {pages} {013082} (\bibinfo {year}
		{2022})}\BibitemShut {NoStop}%
	\bibitem [{\citenamefont {Goto}(2016)}]{goto2016}%
	\BibitemOpen
	\bibinfo {author} {H.~Goto},\ \emph {\bibinfo {title} {Universal Quantum
			Computation with a Nonlinear Oscillator Network}},\ \href
	{\doibase10.1103/PhysRevA.93.050301} {\bibfield  {journal} {\bibinfo
			{journal} {Physical Review A}\ }\textbf {\bibinfo {volume} {93}},\ \bibinfo
		{pages} {050301} (\bibinfo {year} {2016})}\BibitemShut {NoStop}%
	\bibitem [{\citenamefont {Cochrane}\ \emph {et~al.}(1999)\citenamefont
		{Cochrane}, \citenamefont {Milburn},\ and\ \citenamefont
		{Munro}}]{cochrane1999}%
	\BibitemOpen
	\bibinfo {author} {P.~T. Cochrane}, \bibinfo {author} {G.~J. Milburn}\ and\
	\bibinfo {author} {W.~J. Munro},\ \emph {\bibinfo {title} {Macroscopically
			Distinct Quantum-Superposition States as a Bosonic Code for Amplitude
			Damping}},\ \href {\doibase10.1103/PhysRevA.59.2631} {\bibfield  {journal}
		{\bibinfo  {journal} {Physical Review A}\ }\textbf {\bibinfo {volume} {59}},\
		\bibinfo {pages} {2631} (\bibinfo {year} {1999})}\BibitemShut {NoStop}%
	\bibitem [{\citenamefont {Albert}\ \emph {et~al.}(2016)\citenamefont {Albert},
		\citenamefont {Bradlyn}, \citenamefont {Fraas},\ and\ \citenamefont
		{Jiang}}]{albert2016}%
	\BibitemOpen
	\bibinfo {author} {V.~V. Albert}, \bibinfo {author} {B.~Bradlyn}, \bibinfo
	{author} {M.~Fraas}\ and\ \bibinfo {author} {L.~Jiang},\ \emph {\bibinfo
		{title} {Geometry and {{Response}} of {{Lindbladians}}}},\ \href
	{\doibase10.1103/PhysRevX.6.041031} {\bibfield  {journal} {\bibinfo
			{journal} {Physical Review X}\ }\textbf {\bibinfo {volume} {6}},\ \bibinfo
		{pages} {041031} (\bibinfo {year} {2016})}\BibitemShut {NoStop}%
	\bibitem [{\citenamefont {Albert}\ \emph {et~al.}(2018)\citenamefont {Albert},
		\citenamefont {Noh}, \citenamefont {Duivenvoorden}, \citenamefont {Young},
		\citenamefont {Brierley}, \citenamefont {Reinhold}, \citenamefont {Vuillot},
		\citenamefont {Li}, \citenamefont {Shen}, \citenamefont {Girvin},
		\citenamefont {Terhal},\ and\ \citenamefont {Jiang}}]{albert2018a}%
	\BibitemOpen
	\bibinfo {author} {V.~V. Albert}, \bibinfo {author} {K.~Noh}, \bibinfo
	{author} {K.~Duivenvoorden}, \bibinfo {author} {D.~J. Young}, \bibinfo
	{author} {R.~T. Brierley}, \bibinfo {author} {P.~Reinhold}, \bibinfo {author}
	{C.~Vuillot}, \bibinfo {author} {L.~Li}, \bibinfo {author} {C.~Shen},
	\bibinfo {author} {S.~M. Girvin}, \bibinfo {author} {B.~M. Terhal}\ and\
	\bibinfo {author} {L.~Jiang},\ \emph {\bibinfo {title} {Performance and
			Structure of Single-Mode Bosonic Codes}},\ \href
	{\doibase10.1103/PhysRevA.97.032346} {\bibfield  {journal} {\bibinfo
			{journal} {Physical Review A}\ }\textbf {\bibinfo {volume} {97}},\ \bibinfo
		{pages} {032346} (\bibinfo {year} {2018})}\BibitemShut {NoStop}%
	\bibitem [{\citenamefont {Gautier}\ \emph {et~al.}(2022)\citenamefont
		{Gautier}, \citenamefont {Sarlette},\ and\ \citenamefont
		{Mirrahimi}}]{gautier2022}%
	\BibitemOpen
	\bibinfo {author} {R.~Gautier}, \bibinfo {author} {A.~Sarlette}\ and\
	\bibinfo {author} {M.~Mirrahimi},\ \emph {\bibinfo {title} {Combined
			{{Dissipative}} and {{Hamiltonian Confinement}} of {{Cat Qubits}}}},\ \href
	{\doibase10.1103/PRXQuantum.3.020339} {\bibfield  {journal} {\bibinfo
			{journal} {PRX Quantum}\ }\textbf {\bibinfo {volume} {3}},\ \bibinfo {pages}
		{020339} (\bibinfo {year} {2022})}\BibitemShut {NoStop}%
	\bibitem [{\citenamefont {Puri}\ \emph {et~al.}(2017)\citenamefont {Puri},
		\citenamefont {Boutin},\ and\ \citenamefont {Blais}}]{puri2017}%
	\BibitemOpen
	\bibinfo {author} {S.~Puri}, \bibinfo {author} {S.~Boutin}\ and\ \bibinfo
	{author} {A.~Blais},\ \emph {\bibinfo {title} {Engineering the Quantum States
			of Light in a {{Kerr-nonlinear}} Resonator by Two-Photon Driving}},\ \href
	{\doibase10.1038/s41534-017-0019-1} {\bibfield  {journal} {\bibinfo
			{journal} {npj Quantum Information}\ }\textbf {\bibinfo {volume} {3}},\
		\bibinfo {pages} {1} (\bibinfo {year} {2017})}\BibitemShut {NoStop}%
	\bibitem [{\citenamefont {Puri}\ \emph {et~al.}(2019)\citenamefont {Puri},
		\citenamefont {Grimm}, \citenamefont {{Campagne-Ibarcq}}, \citenamefont
		{Eickbusch}, \citenamefont {Noh}, \citenamefont {Roberts}, \citenamefont
		{Jiang}, \citenamefont {Mirrahimi}, \citenamefont {Devoret},\ and\
		\citenamefont {Girvin}}]{puri2019}%
	\BibitemOpen
	\bibinfo {author} {S.~Puri}, \bibinfo {author} {A.~Grimm}, \bibinfo {author}
	{P.~{Campagne-Ibarcq}}, \bibinfo {author} {A.~Eickbusch}, \bibinfo {author}
	{K.~Noh}, \bibinfo {author} {G.~Roberts}, \bibinfo {author} {L.~Jiang},
	\bibinfo {author} {M.~Mirrahimi}, \bibinfo {author} {M.~H. Devoret}\ and\
	\bibinfo {author} {S.~M. Girvin},\ \emph {\bibinfo {title} {Stabilized
			{{Cat}} in a {{Driven Nonlinear Cavity}}: {{A Fault-Tolerant Error Syndrome
					Detector}}}},\ \href {\doibase10.1103/PhysRevX.9.041009} {\bibfield
		{journal} {\bibinfo  {journal} {Physical Review X}\ }\textbf {\bibinfo
			{volume} {9}},\ \bibinfo {pages} {041009} (\bibinfo {year}
		{2019})}\BibitemShut {NoStop}%
	\bibitem [{\citenamefont {Puri}\ \emph {et~al.}(2020)\citenamefont {Puri},
		\citenamefont {{St-Jean}}, \citenamefont {Gross}, \citenamefont {Grimm},
		\citenamefont {Frattini}, \citenamefont {Iyer}, \citenamefont {Krishna},
		\citenamefont {Touzard}, \citenamefont {Jiang}, \citenamefont {Blais},
		\citenamefont {Flammia},\ and\ \citenamefont {Girvin}}]{puri2020}%
	\BibitemOpen
	\bibinfo {author} {S.~Puri}, \bibinfo {author} {L.~{St-Jean}}, \bibinfo
	{author} {J.~A. Gross}, \bibinfo {author} {A.~Grimm}, \bibinfo {author}
	{N.~E. Frattini}, \bibinfo {author} {P.~S. Iyer}, \bibinfo {author}
	{A.~Krishna}, \bibinfo {author} {S.~Touzard}, \bibinfo {author} {L.~Jiang},
	\bibinfo {author} {A.~Blais}, \bibinfo {author} {S.~T. Flammia}\ and\
	\bibinfo {author} {S.~M. Girvin},\ \emph {\bibinfo {title} {Bias-Preserving
			Gates with Stabilized Cat Qubits}},\ \href {\doibase10.1126/sciadv.aay5901}
	{\bibfield  {journal} {\bibinfo  {journal} {Science Advances}\ }\textbf
		{\bibinfo {volume} {6}},\ \bibinfo {pages} {eaay5901} (\bibinfo {year}
		{2020})}\BibitemShut {NoStop}%
	\bibitem [{\citenamefont {Grimm}\ \emph {et~al.}(2020)\citenamefont {Grimm},
		\citenamefont {Frattini}, \citenamefont {Puri}, \citenamefont {Mundhada},
		\citenamefont {Touzard}, \citenamefont {Mirrahimi}, \citenamefont {Girvin},
		\citenamefont {Shankar},\ and\ \citenamefont {Devoret}}]{grimm2020}%
	\BibitemOpen
	\bibinfo {author} {A.~Grimm}, \bibinfo {author} {N.~E. Frattini}, \bibinfo
	{author} {S.~Puri}, \bibinfo {author} {S.~O. Mundhada}, \bibinfo {author}
	{S.~Touzard}, \bibinfo {author} {M.~Mirrahimi}, \bibinfo {author} {S.~M.
		Girvin}, \bibinfo {author} {S.~Shankar}\ and\ \bibinfo {author} {M.~H.
		Devoret},\ \emph {\bibinfo {title} {Stabilization and Operation of a
			{{Kerr-cat}} Qubit}},\ \href {\doibase10.1038/s41586-020-2587-z} {\bibfield
		{journal} {\bibinfo  {journal} {Nature}\ }\textbf {\bibinfo {volume} {584}},\
		\bibinfo {pages} {205} (\bibinfo {year} {2020})}\BibitemShut {NoStop}%
	\bibitem [{\citenamefont {Ruiz}\ \emph {et~al.}(2022)\citenamefont {Ruiz},
		\citenamefont {Gautier}, \citenamefont {Guillaud},\ and\ \citenamefont
		{Mirrahimi}}]{ruiz2022}%
	\BibitemOpen
	\bibinfo {author} {D.~Ruiz}, \bibinfo {author} {R.~Gautier}, \bibinfo
	{author} {J.~Guillaud}\ and\ \bibinfo {author} {M.~Mirrahimi},\ \href@noop {}
	{\emph {\bibinfo {title} {Two-Photon Driven {{Kerr}} Quantum Oscillator with
				Multiple Spectral Degeneracies}}} (\bibinfo {year} {2022}),\ \Eprint
	{http://arxiv.org/abs/2211.03689} {arXiv:2211.03689} \BibitemShut {NoStop}%
	\bibitem [{\citenamefont {Frattini}\ \emph {et~al.}(2022)\citenamefont
		{Frattini}, \citenamefont {Corti{\~n}as}, \citenamefont {Venkatraman},
		\citenamefont {Xiao}, \citenamefont {Su}, \citenamefont {Lei}, \citenamefont
		{Chapman}, \citenamefont {Joshi}, \citenamefont {Girvin}, \citenamefont
		{Schoelkopf}, \citenamefont {Puri},\ and\ \citenamefont
		{Devoret}}]{frattini2022}%
	\BibitemOpen
	\bibinfo {author} {N.~E. Frattini}, \bibinfo {author} {R.~G. Corti{\~n}as},
	\bibinfo {author} {J.~Venkatraman}, \bibinfo {author} {X.~Xiao}, \bibinfo
	{author} {Q.~Su}, \bibinfo {author} {C.~U. Lei}, \bibinfo {author} {B.~J.
		Chapman}, \bibinfo {author} {V.~R. Joshi}, \bibinfo {author} {S.~M. Girvin},
	\bibinfo {author} {R.~J. Schoelkopf}, \bibinfo {author} {S.~Puri}\ and\
	\bibinfo {author} {M.~H. Devoret},\ \href@noop {} {\emph {\bibinfo {title}
			{The Squeezed {{Kerr}} Oscillator: Spectral Kissing and Phase-Flip
				Robustness}}} (\bibinfo {year} {2022}),\ \Eprint
	{http://arxiv.org/abs/2209.03934} {arXiv:2209.03934} \BibitemShut {NoStop}%
	\bibitem [{\citenamefont {Venkatraman}\ \emph {et~al.}(2022)\citenamefont
		{Venkatraman}, \citenamefont {Cortinas}, \citenamefont {Frattini},
		\citenamefont {Xiao},\ and\ \citenamefont {Devoret}}]{Venkatraman2022}%
	\BibitemOpen
	\bibinfo {author} {J.~Venkatraman}, \bibinfo {author} {R.~G. Cortinas},
	\bibinfo {author} {N.~E. Frattini}, \bibinfo {author} {X.~Xiao}\ and\
	\bibinfo {author} {M.~H. Devoret},\ \href@noop {} {\emph {\bibinfo {title}
			{Quantum Interference of Tunneling Paths under a Double-Well Barrier}}}
	(\bibinfo {year} {2022}),\ \Eprint {http://arxiv.org/abs/2211.04605}
	{arXiv:2211.04605} \BibitemShut {NoStop}%
	\bibitem [{\citenamefont {Lieu}\ \emph {et~al.}(2020)\citenamefont {Lieu},
		\citenamefont {Belyansky}, \citenamefont {Young}, \citenamefont {Lundgren},
		\citenamefont {Albert},\ and\ \citenamefont {Gorshkov}}]{lieu2020}%
	\BibitemOpen
	\bibinfo {author} {S.~Lieu}, \bibinfo {author} {R.~Belyansky}, \bibinfo
	{author} {J.~T. Young}, \bibinfo {author} {R.~Lundgren}, \bibinfo {author}
	{V.~V. Albert}\ and\ \bibinfo {author} {A.~V. Gorshkov},\ \emph {\bibinfo
		{title} {Symmetry {{Breaking}} and {{Error Correction}} in {{Open Quantum
					Systems}}}},\ \href {\doibase10.1103/PhysRevLett.125.240405} {\bibfield
		{journal} {\bibinfo  {journal} {Physical Review Letters}\ }\textbf {\bibinfo
			{volume} {125}},\ \bibinfo {pages} {240405} (\bibinfo {year}
		{2020})}\BibitemShut {NoStop}%
	\bibitem [{\citenamefont {Chamberland}\ \emph {et~al.}(2022)\citenamefont
		{Chamberland}, \citenamefont {Noh}, \citenamefont {{Arrangoiz-Arriola}},
		\citenamefont {Campbell}, \citenamefont {Hann}, \citenamefont {Iverson},
		\citenamefont {Putterman}, \citenamefont {Bohdanowicz}, \citenamefont
		{Flammia}, \citenamefont {Keller}, \citenamefont {Refael}, \citenamefont
		{Preskill}, \citenamefont {Jiang}, \citenamefont {{Safavi-Naeini}},
		\citenamefont {Painter},\ and\ \citenamefont
		{Brand{\~a}o}}]{chamberland2022}%
	\BibitemOpen
	\bibinfo {author} {C.~Chamberland}, \bibinfo {author} {K.~Noh}, \bibinfo
	{author} {P.~{Arrangoiz-Arriola}}, \bibinfo {author} {E.~T. Campbell},
	\bibinfo {author} {C.~T. Hann}, \bibinfo {author} {J.~Iverson}, \bibinfo
	{author} {H.~Putterman}, \bibinfo {author} {T.~C. Bohdanowicz}, \bibinfo
	{author} {S.~T. Flammia}, \bibinfo {author} {A.~Keller}, \bibinfo {author}
	{G.~Refael}, \bibinfo {author} {J.~Preskill}, \bibinfo {author} {L.~Jiang},
	\bibinfo {author} {A.~H. {Safavi-Naeini}}, \bibinfo {author} {O.~Painter}\
	and\ \bibinfo {author} {F.~G. Brand{\~a}o},\ \emph {\bibinfo {title}
		{Building a {{Fault-Tolerant Quantum Computer Using Concatenated Cat
					Codes}}}},\ \href {\doibase10.1103/PRXQuantum.3.010329} {\bibfield  {journal}
		{\bibinfo  {journal} {PRX Quantum}\ }\textbf {\bibinfo {volume} {3}},\
		\bibinfo {pages} {010329} (\bibinfo {year} {2022})}\BibitemShut {NoStop}%
	\bibitem [{\citenamefont {Gerry}\ and\ \citenamefont
		{Knight}(2004)}]{gerry2004}%
	\BibitemOpen
	\bibinfo {author} {C.~Gerry}\ and\ \bibinfo {author} {P.~Knight},\ \href
	{\doibase10.1017/CBO9780511791239} {\emph {\bibinfo {title} {Introductory
				{{Quantum Optics}}}}}\ (\bibinfo  {publisher} {{Cambridge University
			Press}},\ \bibinfo {address} {{Cambridge}},\ \bibinfo {year}
	{2004})\BibitemShut {NoStop}%
	\bibitem [{Note1()}]{Note1}%
	\BibitemOpen
	\bibinfo {note} {Note that this choice is arbitrary. An equivalently popular
		convention defines the logical basis as $ \ket {0_L} = \ket {\protect
			\mathcal {C}_\alpha ^+}\protect \tmspace +\thinmuskip {.1667em},\hskip
		1em\relax \ket {1_L} = \ket {\protect \mathcal {C}_\alpha ^-} $. This amounts
		to a $\pi /2$ rotation within the logical space and an exchange between bit
		and phase-flip errors.}\BibitemShut {Stop}%
	\bibitem [{\citenamefont {Hach~III}\ and\ \citenamefont
		{Gerry}(1994)}]{hachiii1994}%
	\BibitemOpen
	\bibinfo {author} {E.~E. Hach~III}\ and\ \bibinfo {author} {C.~C. Gerry},\
	\emph {\bibinfo {title} {Generation of Mixtures of {{Schr\"odinger-cat}}
			States from a Competitive Two-Photon Process}},\ \href
	{\doibase10.1103/PhysRevA.49.490} {\bibfield  {journal} {\bibinfo  {journal}
			{Physical Review A}\ }\textbf {\bibinfo {volume} {49}},\ \bibinfo {pages}
		{490} (\bibinfo {year} {1994})}\BibitemShut {NoStop}%
	\bibitem [{\citenamefont {Gorini}\ \emph {et~al.}(1976)\citenamefont {Gorini},
		\citenamefont {Kossakowski},\ and\ \citenamefont {Sudarshan}}]{gorini1976}%
	\BibitemOpen
	\bibinfo {author} {V.~Gorini}, \bibinfo {author} {A.~Kossakowski}\ and\
	\bibinfo {author} {E.~C.~G. Sudarshan},\ \emph {\bibinfo {title} {Completely
			Positive Dynamical Semigroups of {{N}}-level Systems}},\ \href
	{\doibase10.1063/1.522979} {\bibfield  {journal} {\bibinfo  {journal}
			{Journal of Mathematical Physics}\ }\textbf {\bibinfo {volume} {17}},\
		\bibinfo {pages} {821} (\bibinfo {year} {1976})}\BibitemShut {NoStop}%
	\bibitem [{\citenamefont {Gorini}\ \emph {et~al.}(1978)\citenamefont {Gorini},
		\citenamefont {Frigerio}, \citenamefont {Verri}, \citenamefont
		{Kossakowski},\ and\ \citenamefont {Sudarshan}}]{gorini1978}%
	\BibitemOpen
	\bibinfo {author} {V.~Gorini}, \bibinfo {author} {A.~Frigerio}, \bibinfo
	{author} {M.~Verri}, \bibinfo {author} {A.~Kossakowski}\ and\ \bibinfo
	{author} {E.~C.~G. Sudarshan},\ \emph {\bibinfo {title} {Properties of
			Quantum {{Markovian}} Master Equations}},\ \href
	{\doibase10.1016/0034-4877(78)90050-2} {\bibfield  {journal} {\bibinfo
			{journal} {Reports on Mathematical Physics}\ }\textbf {\bibinfo {volume}
			{13}},\ \bibinfo {pages} {149} (\bibinfo {year} {1978})}\BibitemShut
	{NoStop}%
	\bibitem [{\citenamefont {Lindblad}(1976)}]{lindblad1976}%
	\BibitemOpen
	\bibinfo {author} {G.~Lindblad},\ \emph {\bibinfo {title} {On the Generators
			of Quantum Dynamical Semigroups}},\ \href {\doibase10.1007/BF01608499}
	{\bibfield  {journal} {\bibinfo  {journal} {Communications in Mathematical
				Physics}\ }\textbf {\bibinfo {volume} {48}},\ \bibinfo {pages} {119}
		(\bibinfo {year} {1976})}\BibitemShut {NoStop}%
	\bibitem [{\citenamefont {Lescanne}\ \emph {et~al.}(2020)\citenamefont
		{Lescanne}, \citenamefont {Villiers}, \citenamefont {Peronnin}, \citenamefont
		{Sarlette}, \citenamefont {Delbecq}, \citenamefont {Huard}, \citenamefont
		{Kontos}, \citenamefont {Mirrahimi},\ and\ \citenamefont
		{Leghtas}}]{lescanne2020}%
	\BibitemOpen
	\bibinfo {author} {R.~Lescanne}, \bibinfo {author} {M.~Villiers}, \bibinfo
	{author} {T.~Peronnin}, \bibinfo {author} {A.~Sarlette}, \bibinfo {author}
	{M.~Delbecq}, \bibinfo {author} {B.~Huard}, \bibinfo {author} {T.~Kontos},
	\bibinfo {author} {M.~Mirrahimi}\ and\ \bibinfo {author} {Z.~Leghtas},\ \emph
	{\bibinfo {title} {Exponential Suppression of Bit-Flips in a Qubit Encoded in
			an Oscillator}},\ \href {\doibase10.1038/s41567-020-0824-x} {\bibfield
		{journal} {\bibinfo  {journal} {Nature Physics}\ }\textbf {\bibinfo {volume}
			{16}},\ \bibinfo {pages} {509} (\bibinfo {year} {2020})}\BibitemShut
	{NoStop}%
	\bibitem [{\citenamefont {Guillaud}\ and\ \citenamefont
		{Mirrahimi}(2019)}]{guillaud2019}%
	\BibitemOpen
	\bibinfo {author} {J.~Guillaud}\ and\ \bibinfo {author} {M.~Mirrahimi},\
	\emph {\bibinfo {title} {Repetition {{Cat Qubits}} for {{Fault-Tolerant
					Quantum Computation}}}},\ \href {\doibase10.1103/PhysRevX.9.041053}
	{\bibfield  {journal} {\bibinfo  {journal} {Physical Review X}\ }\textbf
		{\bibinfo {volume} {9}},\ \bibinfo {pages} {041053} (\bibinfo {year}
		{2019})}\BibitemShut {NoStop}%
	\bibitem [{\citenamefont {Guillaud}\ and\ \citenamefont
		{Mirrahimi}(2021)}]{guillaud2021}%
	\BibitemOpen
	\bibinfo {author} {J.~Guillaud}\ and\ \bibinfo {author} {M.~Mirrahimi},\
	\emph {\bibinfo {title} {Error Rates and Resource Overheads of Repetition Cat
			Qubits}},\ \href {\doibase10.1103/PhysRevA.103.042413} {\bibfield  {journal}
		{\bibinfo  {journal} {Physical Review A}\ }\textbf {\bibinfo {volume}
			{103}},\ \bibinfo {pages} {042413} (\bibinfo {year} {2021})}\BibitemShut
	{NoStop}%
	\bibitem [{\citenamefont {{Blume-Kohout}}\ \emph {et~al.}(2010)\citenamefont
		{{Blume-Kohout}}, \citenamefont {Ng}, \citenamefont {Poulin},\ and\
		\citenamefont {Viola}}]{blume-kohout2010}%
	\BibitemOpen
	\bibinfo {author} {R.~{Blume-Kohout}}, \bibinfo {author} {H.~K. Ng}, \bibinfo
	{author} {D.~Poulin}\ and\ \bibinfo {author} {L.~Viola},\ \emph {\bibinfo
		{title} {Information-Preserving Structures: {{A}} General Framework for
			Quantum Zero-Error Information}},\ \href {\doibase10.1103/PhysRevA.82.062306}
	{\bibfield  {journal} {\bibinfo  {journal} {Physical Review A}\ }\textbf
		{\bibinfo {volume} {82}},\ \bibinfo {pages} {062306} (\bibinfo {year}
		{2010})}\BibitemShut {NoStop}%
	\bibitem [{\citenamefont {Bu{\v c}a}\ and\ \citenamefont
		{Prosen}(2012)}]{buca2012}%
	\BibitemOpen
	\bibinfo {author} {B.~Bu{\v c}a}\ and\ \bibinfo {author} {T.~Prosen},\ \emph
	{\bibinfo {title} {A Note on Symmetry Reductions of the {{Lindblad}}
			Equation: Transport in Constrained Open Spin Chains}},\ \href
	{\doibase10.1088/1367-2630/14/7/073007} {\bibfield  {journal} {\bibinfo
			{journal} {New Journal of Physics}\ }\textbf {\bibinfo {volume} {14}},\
		\bibinfo {pages} {073007} (\bibinfo {year} {2012})}\BibitemShut {NoStop}%
	\bibitem [{\citenamefont {S{\'a}nchez~Mu{\~n}oz}\ \emph
		{et~al.}(2018)\citenamefont {S{\'a}nchez~Mu{\~n}oz}, \citenamefont {Lara},
		\citenamefont {Puebla},\ and\ \citenamefont {Nori}}]{sanchezmunoz2018}%
	\BibitemOpen
	\bibinfo {author} {C.~S{\'a}nchez~Mu{\~n}oz}, \bibinfo {author} {A.~Lara},
	\bibinfo {author} {J.~Puebla}\ and\ \bibinfo {author} {F.~Nori},\ \emph
	{\bibinfo {title} {Hybrid {{Systems}} for the {{Generation}} of
			{{Nonclassical Mechanical States}} via {{Quadratic Interactions}}}},\ \href
	{\doibase10.1103/PhysRevLett.121.123604} {\bibfield  {journal} {\bibinfo
			{journal} {Physical Review Letters}\ }\textbf {\bibinfo {volume} {121}},\
		\bibinfo {pages} {123604} (\bibinfo {year} {2018})}\BibitemShut {NoStop}%
	\bibitem [{\citenamefont {S{\'a}nchez~Mu{\~n}oz}\ \emph
		{et~al.}(2019)\citenamefont {S{\'a}nchez~Mu{\~n}oz}, \citenamefont {Bu{\v
				c}a}, \citenamefont {Tindall}, \citenamefont {{Gonz{\'a}lez-Tudela}},
		\citenamefont {Jaksch},\ and\ \citenamefont {Porras}}]{sanchezmunoz2019}%
	\BibitemOpen
	\bibinfo {author} {C.~S{\'a}nchez~Mu{\~n}oz}, \bibinfo {author} {B.~Bu{\v
			c}a}, \bibinfo {author} {J.~Tindall}, \bibinfo {author}
	{A.~{Gonz{\'a}lez-Tudela}}, \bibinfo {author} {D.~Jaksch}\ and\ \bibinfo
	{author} {D.~Porras},\ \emph {\bibinfo {title} {Symmetries and Conservation
			Laws in Quantum Trajectories: {{Dissipative}} Freezing}},\ \href
	{\doibase10.1103/PhysRevA.100.042113} {\bibfield  {journal} {\bibinfo
			{journal} {Physical Review A}\ }\textbf {\bibinfo {volume} {100}},\ \bibinfo
		{pages} {042113} (\bibinfo {year} {2019})}\BibitemShut {NoStop}%
	\bibitem [{\citenamefont {Thingna}\ and\ \citenamefont
		{Manzano}(2021)}]{thingna2021}%
	\BibitemOpen
	\bibinfo {author} {J.~Thingna}\ and\ \bibinfo {author} {D.~Manzano},\ \emph
	{\bibinfo {title} {Degenerated {{Liouvillians}} and Steady-State Reduced
			Density Matrices}},\ \href {\doibase10.1063/5.0045308} {\bibfield  {journal}
		{\bibinfo  {journal} {Chaos: An Interdisciplinary Journal of Nonlinear
				Science}\ }\textbf {\bibinfo {volume} {31}},\ \bibinfo {pages} {073114}
		(\bibinfo {year} {2021})}\BibitemShut {NoStop}%
	\bibitem [{\citenamefont {Zhang}\ \emph {et~al.}(2020)\citenamefont {Zhang},
		\citenamefont {Tindall}, \citenamefont {{Mur-Petit}}, \citenamefont
		{Jaksch},\ and\ \citenamefont {Bu{\v c}a}}]{zhang2020}%
	\BibitemOpen
	\bibinfo {author} {Z.~Zhang}, \bibinfo {author} {J.~Tindall}, \bibinfo
	{author} {J.~{Mur-Petit}}, \bibinfo {author} {D.~Jaksch}\ and\ \bibinfo
	{author} {B.~Bu{\v c}a},\ \emph {\bibinfo {title} {Stationary State
			Degeneracy of Open Quantum Systems with Non-Abelian Symmetries}},\ \href
	{\doibase10.1088/1751-8121/ab88e3} {\bibfield  {journal} {\bibinfo  {journal}
			{Journal of Physics A: Mathematical and Theoretical}\ }\textbf {\bibinfo
			{volume} {53}},\ \bibinfo {pages} {215304} (\bibinfo {year}
		{2020})}\BibitemShut {NoStop}%
	\bibitem [{\citenamefont {Albert}\ and\ \citenamefont
		{Jiang}(2014)}]{albert2014}%
	\BibitemOpen
	\bibinfo {author} {V.~V. Albert}\ and\ \bibinfo {author} {L.~Jiang},\ \emph
	{\bibinfo {title} {Symmetries and Conserved Quantities in {{Lindblad}} Master
			Equations}},\ \href {\doibase10.1103/PhysRevA.89.022118} {\bibfield
		{journal} {\bibinfo  {journal} {Physical Review A}\ }\textbf {\bibinfo
			{volume} {89}},\ \bibinfo {pages} {022118} (\bibinfo {year}
		{2014})}\BibitemShut {NoStop}%
	\bibitem [{\citenamefont {Bu{\v c}a}\ \emph {et~al.}(2019)\citenamefont {Bu{\v
				c}a}, \citenamefont {Tindall},\ and\ \citenamefont {Jaksch}}]{buca2019}%
	\BibitemOpen
	\bibinfo {author} {B.~Bu{\v c}a}, \bibinfo {author} {J.~Tindall}\ and\
	\bibinfo {author} {D.~Jaksch},\ \emph {\bibinfo {title} {Non-Stationary
			Coherent Quantum Many-Body Dynamics through Dissipation}},\ \href
	{\doibase10.1038/s41467-019-09757-y} {\bibfield  {journal} {\bibinfo
			{journal} {Nature Communications}\ }\textbf {\bibinfo {volume} {10}},\
		\bibinfo {pages} {1730} (\bibinfo {year} {2019})}\BibitemShut {NoStop}%
	\bibitem [{\citenamefont {Lidar}(2014)}]{lidar2014}%
	\BibitemOpen
	\bibinfo {author} {D.~A. Lidar},\ in\ \href
	{\doibase10.1002/9781118742631.ch11} {\emph {\bibinfo {booktitle} {Quantum
				{{Information}} and {{Computation}} for {{Chemistry}}}}}\ (\bibinfo
	{publisher} {{John Wiley \& Sons, Ltd}},\ \bibinfo {year} {2014})\ pp.\
	\bibinfo {pages} {295--354}\BibitemShut {NoStop}%
	\bibitem [{\citenamefont {Kempe}\ \emph {et~al.}(2001)\citenamefont {Kempe},
		\citenamefont {Bacon}, \citenamefont {Lidar},\ and\ \citenamefont
		{Whaley}}]{kempe2001}%
	\BibitemOpen
	\bibinfo {author} {J.~Kempe}, \bibinfo {author} {D.~Bacon}, \bibinfo {author}
	{D.~A. Lidar}\ and\ \bibinfo {author} {K.~B. Whaley},\ \emph {\bibinfo
		{title} {Theory of Decoherence-Free Fault-Tolerant Universal Quantum
			Computation}},\ \href {\doibase10.1103/PhysRevA.63.042307} {\bibfield
		{journal} {\bibinfo  {journal} {Physical Review A}\ }\textbf {\bibinfo
			{volume} {63}},\ \bibinfo {pages} {042307} (\bibinfo {year}
		{2001})}\BibitemShut {NoStop}%
	\bibitem [{\citenamefont {Lidar}\ \emph {et~al.}(1998)\citenamefont {Lidar},
		\citenamefont {Chuang},\ and\ \citenamefont {Whaley}}]{lidar1998}%
	\BibitemOpen
	\bibinfo {author} {D.~A. Lidar}, \bibinfo {author} {I.~L. Chuang}\ and\
	\bibinfo {author} {K.~B. Whaley},\ \emph {\bibinfo {title}
		{Decoherence-{{Free Subspaces}} for {{Quantum Computation}}}},\ \href
	{\doibase10.1103/PhysRevLett.81.2594} {\bibfield  {journal} {\bibinfo
			{journal} {Physical Review Letters}\ }\textbf {\bibinfo {volume} {81}},\
		\bibinfo {pages} {2594} (\bibinfo {year} {1998})}\BibitemShut {NoStop}%
	\bibitem [{\citenamefont {Bartolo}\ \emph {et~al.}(2016)\citenamefont
		{Bartolo}, \citenamefont {Minganti}, \citenamefont {Casteels},\ and\
		\citenamefont {Ciuti}}]{bartolo2016}%
	\BibitemOpen
	\bibinfo {author} {N.~Bartolo}, \bibinfo {author} {F.~Minganti}, \bibinfo
	{author} {W.~Casteels}\ and\ \bibinfo {author} {C.~Ciuti},\ \emph {\bibinfo
		{title} {Exact Steady State of a {{Kerr}} Resonator with One- and Two-Photon
			Driving and Dissipation: {{Controllable Wigner-function}} Multimodality and
			Dissipative Phase Transitions}},\ \href {\doibase10.1103/PhysRevA.94.033841}
	{\bibfield  {journal} {\bibinfo  {journal} {Physical Review A}\ }\textbf
		{\bibinfo {volume} {94}},\ \bibinfo {pages} {033841} (\bibinfo {year}
		{2016})}\BibitemShut {NoStop}%
	\bibitem [{\citenamefont {Casteels}\ \emph {et~al.}(2017)\citenamefont
		{Casteels}, \citenamefont {Fazio},\ and\ \citenamefont
		{Ciuti}}]{casteels2017}%
	\BibitemOpen
	\bibinfo {author} {W.~Casteels}, \bibinfo {author} {R.~Fazio}\ and\ \bibinfo
	{author} {C.~Ciuti},\ \emph {\bibinfo {title} {Critical Dynamical Properties
			of a First-Order Dissipative Phase Transition}},\ \href
	{\doibase10.1103/PhysRevA.95.012128} {\bibfield  {journal} {\bibinfo
			{journal} {Physical Review A}\ }\textbf {\bibinfo {volume} {95}},\ \bibinfo
		{pages} {012128} (\bibinfo {year} {2017})}\BibitemShut {NoStop}%
	\bibitem [{\citenamefont {Fink}\ \emph {et~al.}(2018)\citenamefont {Fink},
		\citenamefont {Schade}, \citenamefont {H{\"o}fling}, \citenamefont
		{Schneider},\ and\ \citenamefont {Imamoglu}}]{fink2018}%
	\BibitemOpen
	\bibinfo {author} {T.~Fink}, \bibinfo {author} {A.~Schade}, \bibinfo {author}
	{S.~H{\"o}fling}, \bibinfo {author} {C.~Schneider}\ and\ \bibinfo {author}
	{A.~Imamoglu},\ \emph {\bibinfo {title} {Signatures of a Dissipative Phase
			Transition in Photon Correlation Measurements}},\ \href
	{\doibase10.1038/s41567-017-0020-9} {\bibfield  {journal} {\bibinfo
			{journal} {Nature Physics}\ }\textbf {\bibinfo {volume} {14}},\ \bibinfo
		{pages} {365} (\bibinfo {year} {2018})}\BibitemShut {NoStop}%
	\bibitem [{Note2()}]{Note2}%
	\BibitemOpen
	\bibinfo {note} {At finite driving strength, only the finite-size precursors
		of an actual phase transition are witnessed \cite {minganti2018, rota2017,
			biella2017}. Here, the critical region encompasses a finite range of
		detunings where the system crosses over from the highly-populated phase to
		the other, by showing a series of peaks in the photon number. We define
		$\Delta _c$ as the median value of this critical region.}\BibitemShut {Stop}%
	\bibitem [{\citenamefont {Rivas}\ and\ \citenamefont
		{Huelga}(2012)}]{rivas2012}%
	\BibitemOpen
	\bibinfo {author} {A.~Rivas}\ and\ \bibinfo {author} {S.~F. Huelga},\ \href
	{\doibase10.1007/978-3-642-23354-8} {\emph {\bibinfo {title} {Open {{Quantum
						Systems}}: {{An Introduction}}}}},\ {{SpringerBriefs}} in {{Physics}}\
	(\bibinfo  {publisher} {{Springer}},\ \bibinfo {address} {{Berlin,
			Heidelberg}},\ \bibinfo {year} {2012})\BibitemShut {NoStop}%
	\bibitem [{\citenamefont {Minganti}(2018)}]{minganti2018}%
	\BibitemOpen
	\bibinfo {author} {F.~Minganti},\ \emph {\bibinfo {title}
		{Out-of-{{Equilibrium Phase Transitions}} in {{Nonlinear Optical
					Systems}}}},\ \href@noop {} {Ph.D. thesis},\ \bibinfo  {school} {Universit\'e
		Sorbonne Paris Cit\'e} (\bibinfo {year} {2018})\BibitemShut {NoStop}%
	\bibitem [{\citenamefont {Savona}(2017)}]{savona2017}%
	\BibitemOpen
	\bibinfo {author} {V.~Savona},\ \emph {\bibinfo {title} {Spontaneous Symmetry
			Breaking in a Quadratically Driven Nonlinear Photonic Lattice}},\ \href
	{\doibase10.1103/PhysRevA.96.033826} {\bibfield  {journal} {\bibinfo
			{journal} {Physical Review A}\ }\textbf {\bibinfo {volume} {96}},\ \bibinfo
		{pages} {033826} (\bibinfo {year} {2017})}\BibitemShut {NoStop}%
	\bibitem [{\citenamefont {Di~Candia}\ \emph {et~al.}(2021)\citenamefont
		{Di~Candia}, \citenamefont {Minganti}, \citenamefont {Petrovnin},
		\citenamefont {Paraoanu},\ and\ \citenamefont {Felicetti}}]{dicandia2021}%
	\BibitemOpen
	\bibinfo {author} {R.~Di~Candia}, \bibinfo {author} {F.~Minganti}, \bibinfo
	{author} {K.~V. Petrovnin}, \bibinfo {author} {G.~S. Paraoanu}\ and\ \bibinfo
	{author} {S.~Felicetti},\ \href@noop {} {\emph {\bibinfo {title} {Critical
				Parametric Quantum Sensing}}} (\bibinfo {year} {2021}),\ \Eprint
	{http://arxiv.org/abs/2107.04503} {arXiv:2107.04503} \BibitemShut {NoStop}%
	\bibitem [{\citenamefont {Ilias}\ \emph {et~al.}(2022)\citenamefont {Ilias},
		\citenamefont {Yang}, \citenamefont {Huelga},\ and\ \citenamefont
		{Plenio}}]{ilias2022}%
	\BibitemOpen
	\bibinfo {author} {T.~Ilias}, \bibinfo {author} {D.~Yang}, \bibinfo {author}
	{S.~F. Huelga}\ and\ \bibinfo {author} {M.~B. Plenio},\ \emph {\bibinfo
		{title} {Criticality-{{Enhanced Quantum Sensing}} via {{Continuous
					Measurement}}}},\ \href {\doibase10.1103/PRXQuantum.3.010354} {\bibfield
		{journal} {\bibinfo  {journal} {PRX Quantum}\ }\textbf {\bibinfo {volume}
			{3}},\ \bibinfo {pages} {010354} (\bibinfo {year} {2022})}\BibitemShut
	{NoStop}%
	\bibitem [{\citenamefont {{Fern{\'a}ndez-Lorenzo}}\ and\ \citenamefont
		{Porras}(2017)}]{fernandez-lorenzo2017}%
	\BibitemOpen
	\bibinfo {author} {S.~{Fern{\'a}ndez-Lorenzo}}\ and\ \bibinfo {author}
	{D.~Porras},\ \emph {\bibinfo {title} {Quantum Sensing Close to a Dissipative
			Phase Transition: {{Symmetry}} Breaking and Criticality as Metrological
			Resources}},\ \href {\doibase10.1103/PhysRevA.96.013817} {\bibfield
		{journal} {\bibinfo  {journal} {Physical Review A}\ }\textbf {\bibinfo
			{volume} {96}},\ \bibinfo {pages} {013817} (\bibinfo {year}
		{2017})}\BibitemShut {NoStop}%
	\bibitem [{Note3()}]{Note3}%
	\BibitemOpen
	\bibinfo {note} {We choose the overlap to quantify the resistance of critical
		cat because it is more numerically stable than other indicator (e.g., the
		fidelity). Qualitatively analogous results are found using the fidelity
		measure for the distance between the initial and final states.}\BibitemShut
	{Stop}%
	\bibitem [{\citenamefont {Fortunato}\ \emph {et~al.}(2003)\citenamefont
		{Fortunato}, \citenamefont {Viola}, \citenamefont {Pravia}, \citenamefont
		{Knill}, \citenamefont {Laflamme}, \citenamefont {Havel},\ and\ \citenamefont
		{Cory}}]{fortunato2003}%
	\BibitemOpen
	\bibinfo {author} {E.~M. Fortunato}, \bibinfo {author} {L.~Viola}, \bibinfo
	{author} {M.~A. Pravia}, \bibinfo {author} {E.~Knill}, \bibinfo {author}
	{R.~Laflamme}, \bibinfo {author} {T.~F. Havel}\ and\ \bibinfo {author} {D.~G.
		Cory},\ \emph {\bibinfo {title} {Exploring Noiseless Subsystems via Nuclear
			Magnetic Resonance}},\ \href {\doibase10.1103/PhysRevA.67.062303} {\bibfield
		{journal} {\bibinfo  {journal} {Physical Review A}\ }\textbf {\bibinfo
			{volume} {67}},\ \bibinfo {pages} {062303} (\bibinfo {year}
		{2003})}\BibitemShut {NoStop}%
	\bibitem [{Note4()}]{Note4}%
	\BibitemOpen
	\bibinfo {note} {This is true for all $\protect \mathaccentV {hat}05E\rho
		_j^\mu $ except for $\protect \mathaccentV {hat}05E\rho _0^+ \equiv \protect
		\mathaccentV {hat}05E{\rho }_{\protect \rm ss}$. To simplify notation, we
		will refer to $\protect \mathaccentV {hat}05E\rho _j^\mu $ as any element in
		$\protect \{\protect \mathaccentV {hat}05E\rho _j^\mu \protect \}_{\mu \in
			\protect \{\pm \protect \}, j\in \protect \mathbb {N}}/\protect \{\protect
		\mathaccentV {hat}05E\rho _0^+\protect \}$ and use $\protect \mathaccentV
		{hat}05E{\rho }_{\protect \rm ss}$ to identify $\protect \mathaccentV
		{hat}05E\rho _0^+$.}\BibitemShut {Stop}%
\end{thebibliography}

%

\end{document}